\documentclass[longauth, traditabstract]{aa}
\usepackage{txfonts}
\usepackage{graphicx}
\usepackage{natbib}
\usepackage{xspace}
\usepackage{color}
\bibpunct{(}{)}{;}{a}{}{,}

\def\mpcoh{\,h^{-1}{\rm Mpc}}

\def\kmsmpc{\,{\rm km\,s^{-1}Mpc^{-1}}}

\newcommand{\dif}{{\mathrm d}}
\newcommand{\mm}{{\mathcal M}}

\begin{document}

\title{The VIMOS Public Extragalactic Redshift Survey (VIPERS)
    \thanks{based on observations collected at the European Southern
    Observatory, Cerro Paranal, Chile, using the Very Large Telescope
    under programmes 182.A-0886 and partly 070.A-9007.  Also based on
    observations obtained with MegaPrime/MegaCam, a joint project of
    CFHT and CEA/DAPNIA, at the Canada-France-Hawaii Telescope (CFHT),
    which is operated by the National Research Council (NRC) of
    Canada, the Institut National des Science de l'Univers of the
    Centre National de la Recherche Scientifique (CNRS) of France, and
    the University of Hawaii.  This work is based in part on data
    products produced at TERAPIX and the Canadian Astronomy Data
    Centre as part of the Canada-France-Hawaii Telescope Legacy
    Survey, a collaborative project of NRC and CNRS.  The VIPERS web
    site is http://www.vipers.inaf.it/. }
}

\subtitle{On the correct recovery of the count-in-cell probability
  distribution function.}
%
%
\author{
 J.~Bel\inst{2}
\and E.~Branchini\inst{10,28,29}
\and C.~Di Porto\inst{9}
\and O.~Cucciati\inst{17,9}        
\and B.~R.~Granett\inst{2}
\and A.~Iovino\inst{2}
\and S.~de la Torre\inst{4}
\and C.~Marinoni\inst{7,30,31}
\and L.~Guzzo\inst{2,27}
\and L.~Moscardini\inst{17,18,9}
\and A.~Cappi\inst{9,21}
\and U.~Abbas\inst{5}
\and C.~Adami\inst{4}
\and S.~Arnouts\inst{6}
\and M.~Bolzonella\inst{9}           
\and D.~Bottini\inst{3}
\and J.~Coupon\inst{32}  
\and I.~Davidzon\inst{9,17}
\and G.~De Lucia\inst{13}
\and A.~Fritz\inst{3}
\and P.~Franzetti\inst{3}
\and M.~Fumana\inst{3}
\and B.~Garilli\inst{3,4}     
\and O.~Ilbert\inst{4}
\and J.~Krywult\inst{15}
\and V.~Le Brun\inst{4}
\and O.~Le F\`evre\inst{4}
\and D.~Maccagni\inst{3}
\and K.~Ma{\l}ek\inst{23}
\and F.~Marulli\inst{17,18,9}
\and H.~J.~McCracken\inst{19}
\and L.~Paioro\inst{3}
\and M.~Polletta\inst{3}
\and A.~Pollo\inst{22,23}
\and H.~Schlagenhaufer\inst{24,20}
\and M.~Scodeggio\inst{3} 
\and L.~A.~.M.~Tasca\inst{4}
\and R.~Tojeiro\inst{11}
\and D.~Vergani\inst{25,9}
\and A.~Zanichelli\inst{26}
\and A.~Burden\inst{11}
\and A.~Marchetti\inst{1,2} 
\and Y.~Mellier\inst{19}
\and R.~C.~Nichol\inst{11}
\and J.~A.~Peacock\inst{14}
\and W.~J.~Percival\inst{11}
\and S.~Phleps\inst{20}
\and M.~Wolk\inst{19}
}

\offprints{ Bel., J. \\ \email{julien.bel@brera.inaf.it} }

\institute{
Universit\`{a} degli Studi di Milano, via G. Celoria 16, 20130 Milano, Italy 
\and INAF - Osservatorio Astronomico di Brera, Via Brera 28, 20122 Milano, via E. Bianchi 46, 23807 Merate, Italy 
\and INAF - Istituto di Astrofisica Spaziale e Fisica Cosmica Milano, via Bassini 15, 20133 Milano, Italy
\and Aix Marseille Universit\'e, CNRS, LAM (Laboratoire d'Astrophysique de Marseille) UMR 7326, 13388, Marseille, France  
\and INAF - Osservatorio Astronomico di Torino, 10025 Pino Torinese, Italy 
\and Canada-France-Hawaii Telescope, 65--1238 Mamalahoa Highway, Kamuela, HI 96743, USA 
\and Aix Marseille Universit\'e, CNRS, CPT, UMR 7332, 13288 Marseille, France   
\and Universit\'{e} de Lyon, F-69003 Lyon, France 
\and INAF - Osservatorio Astronomico di Bologna, via Ranzani 1, I-40127, Bologna, Italy 
\and Dipartimento di Matematica e Fisica, Universit\`{a} degli Studi Roma Tre, via della Vasca Navale 84, 00146 Roma, Italy 
\and Institute of Cosmology and Gravitation, Dennis Sciama Building, University of Portsmouth, Burnaby Road, Portsmouth, PO1 3FX 
\and Institute of Astronomy and Astrophysics, Academia Sinica, P.O. Box 23-141, Taipei 10617, Taiwan
\and INAF - Osservatorio Astronomico di Trieste, via G. B. Tiepolo 11, 34143 Trieste, Italy 
\and SUPA, Institute for Astronomy, University of Edinburgh, Royal Observatory, Blackford Hill, Edinburgh EH9 3HJ, UK 
\and Institute of Physics, Jan Kochanowski University, ul. Swietokrzyska 15, 25-406 Kielce, Poland 
\and Department of Particle and Astrophysical Science, Nagoya University, Furo-cho, Chikusa-ku, 464-8602 Nagoya, Japan 
\and Dipartimento di Fisica e Astronomia - Alma Mater Studiorum Universit\`{a} di Bologna, viale Berti Pichat 6/2, I-40127 Bologna, Italy 
\and INFN, Sezione di Bologna, viale Berti Pichat 6/2, I-40127 Bologna, Italy 
\and Institute d'Astrophysique de Paris, UMR7095 CNRS, Universit\'{e} Pierre et Marie Curie, 98 bis Boulevard Arago, 75014 Paris, France 
\and Max-Planck-Institut f\"{u}r Extraterrestrische Physik, D-84571 Garching b. M\"{u}nchen, Germany 
\and Laboratoire Lagrange, UMR7293, Universit\'e de Nice Sophia Antipolis, CNRS, Observatoire de la C\^ote d’Azur, 06300 Nice, France 
\and Astronomical Observatory of the Jagiellonian University, Orla 171, 30-001 Cracow, Poland 
\and National Centre for Nuclear Research, ul. Hoza 69, 00-681 Warszawa, Poland 
\and Universit\"{a}tssternwarte M\"{u}nchen, Ludwig-Maximillians Universit\"{a}t, Scheinerstr. 1, D-81679 M\"{u}nchen, Germany 
\and INAF - Istituto di Astrofisica Spaziale e Fisica Cosmica Bologna, via Gobetti 101, I-40129 Bologna, Italy 
\and INAF - Istituto di Radioastronomia, via Gobetti 101, I-40129, Bologna, Italy 
\and Dipartimento di Fisica, Universit\`a di Milano-Bicocca, P.zza della Scienza 3, I-20126 Milano, Italy 
\and INFN, Sezione di Roma Tre, via della Vasca Navale 84, I-00146 Roma, Italy 
\and INAF - Osservatorio Astronomico di Roma, via Frascati 33, I-00040 Monte Porzio Catone (RM), Italy 
\and Institut Universitaire de France 
\and Universit\'e de Toulon, CNRS, CPT, UMR 7332, 83957 La Garde, France 
\and Astronomical Observatory of the University of Geneva, ch. d’Ecogia 16, CH-1290 Versoix, Switzerland 
}
%
%
\date{Received --; accepted --}
%

\abstract{
We compare three methods to measure the count-in-cell probability density
function of galaxies in a spectroscopic redshift survey.  From this
comparison we found that when the sampling is low (the average number
of object per cell is around unity) it is necessary to use a
parametric method to model the galaxy distribution.  We used a set of
mock catalogues of VIPERS, in order to verify if we were able to
reconstruct the cell-count probability distribution once the
observational strategy is applied. We find that in the simulated catalogues, the probability distribution of galaxies is better represented by a Gamma expansion than a Skewed Log-Normal.   Finally, we correct the cell-count probability
distribution function from the angular selection effect of the VIMOS instrument and study the redshift and absolute magnitude dependency of the underlying galaxy density function in VIPERS from redshift $0.5$ to $1.1$. We found very weak evolution of the probability density distribution function and that it is well approximated, independently from the chosen tracers, by a Gamma distribution. 
}


\maketitle

\section{Introduction}

The galaxy clustering offers a formidable playground to try to understand
how structures have been growing during the evolution of the universe. A number
of statistical tools have been developed and used over the past thirty years \citep[see][for a review]{Bernardeau02}.
In general, these statistical methods use the fact that the clustering of galaxies is due to the gravitational pull of the underlying
matter distribution. Hence, the study of the spatial distribution of galaxies in the universe allows us to get information about
the statistical properties of its matter content. As a result, it is of paramount importance to be able to measure the statistical quantities
describing the galaxy distribution from a redshift survey.

The development of multi-object spectrographs on 8-m class telescopes
during the 1990s triggered a number of deep redshift surveys with
measured distances beyond $z\sim 0.5$ over areas of 1--2 deg$^2$
(e.g. VVDS \citealt{lefevre}, DEEP2 \citealt{newman} and zCOSMOS
\citealt{lilly09}). Even so, it was not until the wide extension of
VVDS was produced \citep{Garilli08}, that a survey existed with
sufficient volume to attempt cosmologically meaningful computations at
$z\sim 1$ \citep{Guzzo}.  In general, clustering measurements at
$z\simeq 1$ from these samples remained dominated by cosmic variance,
as dramatically shown by the discrepancy observed between the VVDS and
zCOSMOS correlation functions at $z\simeq 0.8$ \citep{delaTorre10}.
 
The VIMOS Public Extragalactic Redshift Survey (VIPERS) is part of
this global attempt to take cosmological measurements at $z\sim 1$ to
a new level in terms of statistical significance. In contrast to the
BOSS and WiggleZ surveys, which use large-field-of-view ($\sim 1$
deg$^2$) fibre optic positioners to probe huge volumes at low sampling
density, VIPERS exploits the features of VIMOS at the ESO VLT to yield
a dense galaxy sampling over a moderately large field of view ($\sim
0.08$ deg$^2$). It reaches a volume at $0.5<z<1.2$ comparable to that
of the 2dFGRS \citep{colless} at $z\sim 0.1$, allowing the
cosmological evolution to be tested with small statistical errors.

The VIPERS redshifts are being collected by tiling the
selected sky areas with a uniform
mosaic of VIMOS fields. The area covered is not contiguous, but
presents regular gaps due to the specific footprint of the instrument
field of view, in addition to intrinsic unobserved areas due to bright
stars or defects in the original photometric catalogue.  The VIMOS
field of view has four rectangular
regions of about $8 \times 7$ square arcminutes each, separated by an
unobserved cross \citep{Guzzo13, delaTorre}.
This creates a regular pattern of gaps in the angular
distribution of the measured galaxies. Additionally, the Target Sampling Rate
and the Survey Success Rate vary among the quadrants, and a few of the
latter were lost because of mechanical problems within VIMOS \citep{Garilli14}. 
Finally, the slit-positioning algorithm, SPOC \citep[see][]{Bottini05}, also introduces some
small-scale angular selection effects,   
with different constraints along the dispersion and spatial directions
of the spectra, as thoroughly discussed in \citet{delaTorre}.
Clearly, this combination of angular selection effects has to be taken
properly into account when estimating any clustering statistics.

In this paper we measure the probability distribution function from the VIPERS
Public Data Release 1 (PDR-1) redshift catalogue, including $\sim 64\%$ of the final number of
redshifts expected at completion  (see \citealt{Guzzo13,Garilli14} for a detailed
description of the survey data set).  The paper is organized as follows. In \S 2 we introduce the VIPERS
survey and the features of the PDR-1 sample. In \S 3 we review the basics of the
three methods we compared. In \S 4 we present a null test of the three
method on a synthetic galaxy catalogue. In \S 5
we use galaxy mock catalogues to assess performances of two of the methods. Magnitude and redshift dependance of the
probability distribution function of VIPERS PDR-1 galaxies are presented in  \S 6 and
conclusions are drawn in \S 7. 

Throughout, the Hubble constant is parameterized via $h=H_0/100\kmsmpc$, all magnitudes in this paper are in the AB system \citep{og} and
we will not give an explicit AB suffix. In order to convert redshifts into comoving distances we assume that the matter density parameter is $\Omega_m=0.27$ and that the universe is spatially flat with a $\Lambda$CDM cosmology without radiations.

\section{Data}
\label{data}

%
The VIMOS Public
Extragalactic Redshift Survey (VIPERS) is a spectroscopic redshift survey being built using the
VIMOS spectrograph at the ESO VLT. The survey target sample is
selected from the Canada-France-Hawaii Telescope Legacy Survey Wide
(CFHTLS-Wide) optical photometric catalogues \citep{CFHTLS}. The final VIPERS
will cover $\sim24$ deg$^2$ on the sky, divided over two areas within the
W1 and W4 CFHTLS fields. Galaxies are selected to a limit of
$i_{AB}<22.5$, further applying a simple and robust $gri$ colour
pre-selection, as to effectively remove galaxies at $z<0.5$. Coupled
to an aggressive observing strategy \citep{Scodeggio09}, this
allows us to double the galaxy sampling rate in the redshift range of
interest, with respect to a pure magnitude-limited sample ($\sim
40\%$). At the same time, the area and depth of the survey result in a
fairly large volume, $\sim 5 \times 10^{7}$ h$^{-3}$ Mpc$^{3}$,
analogous to that of the 2dFGRS at $z\sim 0.1$. Such combination of
sampling and depth is quite unique over current redshift surveys at
$z>0.5$. The VIPERS spectra are collected with the VIMOS multi-object
spectrograph \citep{vimos_ref} at moderate resolution ($R=210$), using
the LR Red grism, providing a wavelength coverage of
5500-9500${\rm\AA}$ and a typical redshift error of $141(1+z)$ km
sec$^{-1}$ . The full VIPERS area is covered
through a mosaic of 288 VIMOS pointings (192 in the W1 area, and 96 in
the W4 area). A discussion of the survey data reduction and management
infrastructure is presented in \citet{Garilli2012}. An early subset of
the spectra used here is analyzed and classified through a Principal
Component Analysis (PCA) in \citet{Marchetti2012}.

A quality flag is assigned to each measured redshift, based on the
quality of the corresponding spectrum. Here and in all parallel VIPERS
science analyses we use only galaxies with flags 2 to 9 inclusive,
corresponding to a global redshift confidence level of 98\%. The
redshift confirmation rate and redshift accuracy have been estimated
using repeated spectroscopic observations in the VIPERS fields. A more complete description of the survey
construction, from the definition of the target sample to the actual
spectra and redshift measurements, is given in the parallel survey
description paper \citep{Guzzo13}.

\begin{figure*}
\centerline{\includegraphics[width=180mm,angle=0]{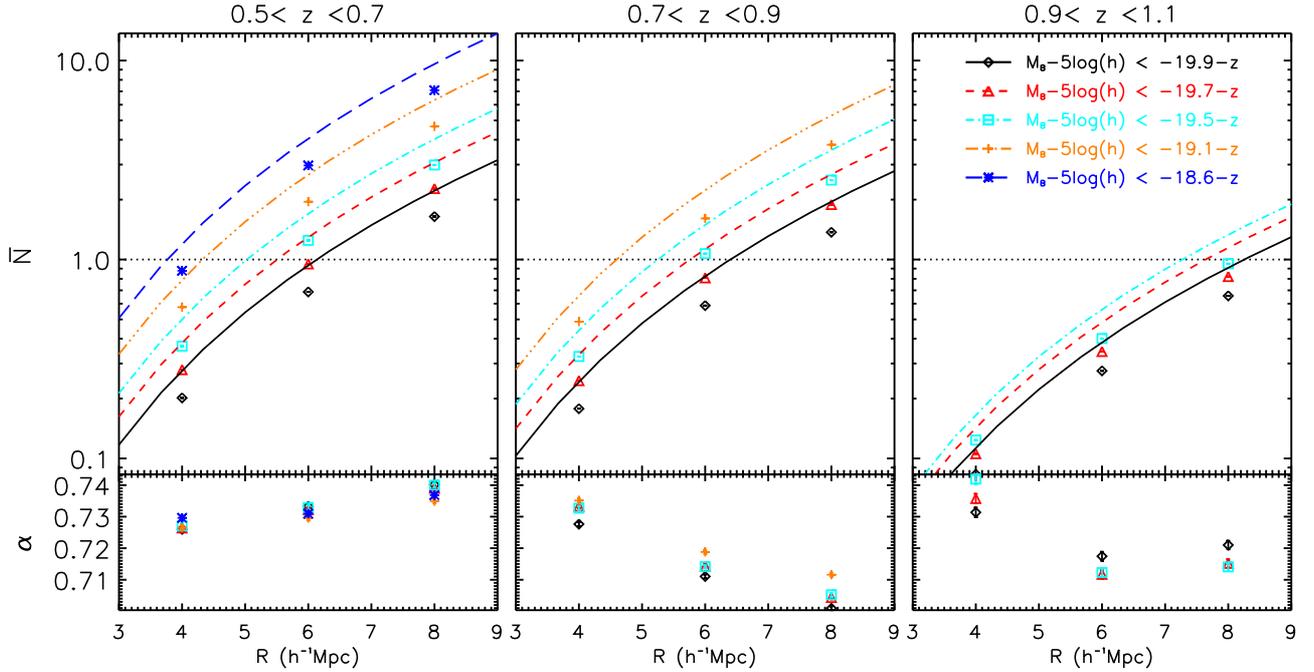}}
\caption{\small \textit{Upper}: Expected mean number count in spheres (solid line, from Eq.~\ref{nbarth}) with respect to the observed one (symbols) for the various luminosity cuts and for the three redshift bins $[0.5,0.7]$ (left panel), $[0.7,0.9]$ (central panel), and $[0.9,1.1]$ (right panel). The selection in absolute magnitude $M_B$ in B-band corresponding to each symbols/lines and colors are indicated in the inset. The dotted line displays the $\bar N=1$. \textit{Lower}: Displays the deviation $\alpha$ (see Eq.~\ref{alpha}) between the expected mean number $\bar{\mathcal{N}}_R$ and the observed one $\bar N$ with respect to the radius $R$ of the cells.
 }  
\label{nbar}
\end{figure*}

The data set used in this paper and the other papers of this early science
release is the VIPERS Public Data Release 1 (PDR-1) catalogue, which
have been made publicly available in September 2013.  This includes
$55,359$ objects, spread over a global area of $8.6\times 1.0$ deg$^2$
and $5.3\times 1.5$ deg$^2$ respectively in W1 and W4. It corresponds
to the data frozen in the VIPERS database at the end of the 2011/2012
observing campaign, i.e. 64\% of the final expected survey. 
For the specific analysis presented here, the sample has been
further limited to its higher-redshift part, selecting only galaxies
with $0.55<z< 1.1$. The reason for this selection is related to
minimizing the shot noise and maximizing the volume. 
 This reduces the usable sample to $18135$ and
 $16879$ galaxies in W1 and W4 respectively (always with quality flags between $2$ and $9$).  The corresponding effective volume of the two samples are
$6.57$ and $6.14$ $\times 10^{6}$ h$^{-3}$ Mpc$^{3}$. At redshift $z=1.1$ they  spann respectively the angular comoving distances  $\sim 370$ and $ 230\mpcoh$.  We divide the W1 and W4 fields in three redshift bins and we build magnitude limited sub-samples in each of them. For convenience, we use the magnitude limits listed in Table (1) of  \citet{diPorto}, which we recall in Tab.~(\ref{lumivip}).

\begin{table}
\centering
  \label{lumivip}
\caption{List of the magnitude selected objects (in B-band) in the VIPERS PDR-1}
  \begin{tabular}{c c c c}
 \hline
 $z_{\mathrm{min}}$ & $z_{\mathrm{min}}$ & luminosity & $\bar \rho$ (Eq. \ref{density}) \\
 &   & M$_B - 5\log(h) <$  & $10^{-3}h^3$Mpc$^{-3}$ \\
     \hline 
$0.5$	&	$0.7$	&	$-18.6-z$	&   $4.49$     \\ 
$0.5$	&    $0.7$	&	$-19.1-z$	&   $2.96$     \\ 
$0.5$	&	$0.7$	&	$-19.5-z$	&   $1.88$     \\ 
$0.5$	&	$0.7$	&	$-19.7-z$	&   $1.43$     \\ 
$0.5$	&	$0.7$	&	$-19.9-z$	&   $1.04$     \\ 
\hline
$0.7$	&    $0.9$	&	$-19.1-z$	&   $2.47$    \\ 
$0.7$	&	$0.9$	&	$-19.5-z$	&   $1.66$     \\ 
$0.7$	&	$0.9$	&	$-19.7-z$	&   $1.25$    \\ 
$0.7$	&	$0.9$	&	$-19.9-z$	&   $0.912$    \\ 
\hline
$0.9$	&	$1.1$	&	$-19.5-z$	&   $0.622$    \\ 
$0.9$	&	$1.1$	&	$-19.7-z$	&   $0.535$    \\ 
$0.9$	&	$1.1$	&	$-19.9-z$	&   $0.425$     \\ 
\hline
   \end{tabular}
\end{table}

The VIMOS footprint has an important impact on the observed
probability of finding $N$ galaxies in a randomly placed spherical
cell in the survey volume. As a matter of fact, a direct
appreciation of the masked area can be shown on the first moment of the probability distribution, i.e. the expectation value of the number count $\bar N\equiv \sum_{N=0}^{\infty} N P_N$. On one hand, we can predict the mean number of objects per cells from the knowledge of the number density in each considered redshift bins and on the other hand we can estimate it by placing a regular grid of spherical cells of radius $R$ into the volume surveyed by VIPERS.
In fact, given the solid angle of W1 and W4 and the corresponding number of galaxies $N_1$ and $N_4$ contained in a redshift bin extracted from each field, one can estimate the total number density as

\begin{equation}
\bar\rho=\frac{N_1+N_4}{\Omega_1 + \Omega_4}\frac{1}{V_k},
\label{density}
\end{equation}
where $V_k$ is defined as the volume corresponding to a sector of a spherical shell with solid angle equal to unity. In the case of VIPERS PDR-1 the effective solide angles corresponding to W1 and W4 are respectively $\Omega_1=1.6651683\times 10^{-3}$ and $\Omega_4=1.5573021\times 10^{-3}$ (in square radians). One can therefore predict the corresponding expected number of objects in each cell by multiplying the averaged number density by the volume of a cell. It reads 

\begin{equation}
\bar{\mathcal{N}}_R=\frac{4}{3}\pi R^3 \bar\rho,
\label{nbarth}
\end{equation}
in the case of the spherical cells of radius $R$ considered in this work. The expectation value
$\bar\mathcal{N}_R$ with respect to the radius of the cells
corresponding to each luminosity sub-sample extracted from VIPERS-PDR1 is represented by lines
in Fig.~(\ref{nbar}). On the same figure we display the
measured mean number of object $\bar N$ in each redshift bins. Note that
to perform this measurement we place a grid of equally separated  ($4h^{-1}$Mpc) 
spheres of radius $R={4,6,8}h^{-1}$Mpc and we reject spheres with more than $40$\% of their volume outside the observed region \citep[see][]{bel13}. We quantify the effect of the mask using the quantity 

\begin{equation}
\alpha\equiv \frac{\bar N}{\bar{\mathcal{N}}_R},
\label{alpha}
\end{equation}
in fact the botton panels of Fig.~(\ref{nbar}) shows that for all sub samples and at all redshifts the neat effect of the masks is to under-sample the galaxy field by roughly $72$\%. It also shows that the correction factor $\alpha$ depends on the considered redshift, on the luminosity and on the cell-size. The scale dependency can be explained by the fact that the correction parameter $\alpha$ depends on how the cells overlap with the masked regions. The left panel of Fig.~(\ref{nbar}) suggests that at low redshift the mask effect behaves in the same way for all the luminosity samples while the middle panel shows a clear dependency with respect to luminosity. The correction factor $\alpha$ depends on the redshift distribution, as a result the apparent dependency with respect to the luminosity is due to the dependence of the number density with respect to the luminosity of the considered objects. 


The mask not only modifies the mean number of object but it  also modifies the higher order moments of the distribution, such that the measured $P_N$ will be systematically altered. In the present paper we show that this systematic effect can be taken into account by measuring the underlying probability density function of the galaxy density contrast $\delta$. It has been shown \citep[see Fig.~(8) of][]{bel13} that after rejecting spheres with more than $40$\% of their volume outside the survey, the local poisson process approximation holds. In particular, it allows to use the ``wrong'' probability distribution function in order to get reliable information on the underlying probability density function $p(\delta)$. Then applying the Poisson sampling one can recover the unaltered $P_N$ using that $\bar N=\bar N(masked)/\alpha$. For the sake of completeness we provide the reader with the measured probability function obtained after rejecting the cells with more than $40$\% of their volume outside the survey (see Fig.~\ref{pnmask}).

In particular, let $P_M$ and $P_N$, respectively, be the observed and the true Counting Probability Distribution Function (CPDF). Assuming that from the knowledge of $P_M$ there exists a process to get the underlying probability density function of the stochastic field $\Lambda$, which is associated to the random variable $N$, one can compute the true CPDF applying

\begin{equation}
P_N=\int_0^\infty P[N|\Lambda]p(\Lambda)\dif\Lambda,
\label{sampling}
\end{equation} 
where $P[N|\Lambda]$ is called the sampling conditional probability; it determines the sampling process from which the discrete cell-count arises. In the following we assume that this sampling conditional probability follows a Poisson law \citep{Lay}, as a result in Eq.~(\ref{sampling}) we substitute

\begin{equation}
P[N|\Lambda]=K[N, \Lambda]\equiv \frac{\Lambda^N}{N!}e^{-\Lambda}.
\label{pkernel}
\end{equation}
It is also convenient to express Eq.~(\ref{sampling}) in terms of the density contrast of the stochastic field $\Lambda$, $\delta\equiv \Lambda/\bar\Lambda -1$, it follows that

\begin{equation}
P_N=\int_{-1}^\infty K[N|\bar N(1+\delta)]p(\delta)\dif\delta,
\label{samplingd}
\end{equation} 
where we used that $\bar \Lambda=\bar N$, which is a property of the Poisson sampling.

Continuing along this direction that we propose to compare three methods which aim at extracting the underlying probability density  function (PDF) in order to correct the observed CPDF from the angular selection effects of VIPERS. 

\section{Methods}

 In this section we review the PDF estimators that we use and compare with each others in this paper. The purpose is to select the method which will be more adapted to the VIPERS characteristics.

\subsection{The Richardson-Lucy deconvolution}

This is an iterative method which aims at inverting Eq.~(\ref{samplingd}) without parametrizing the underlying PDF, it has been investigated by \citet{S&P04}. This method starts with an initial guess $p_0$ for the probability density function $p$ which is used to compute the corresponding expected observed $P_{N,0}$ via
$$
P_{N,0}=\int_{-1}^{\infty} \hat K\left[ N, \bar N(1+\delta)\right]p_0(\delta)\dif\delta,
$$
where $\hat K\left[ N, \bar N(1+\delta)\right]\equiv K/\sum_N K$.
The probability density function used at the next step is obtained using
$$
\hat p_{i+1}(\delta)=\hat p_{i}(\delta) \sum_{N=0}^{N_{max}}\frac{P_N}{P_{N,i}} \hat K\left[ N, \bar N(1+\delta)\right],
$$
where $\hat p \equiv p\sum_N K$.
For each step the agreement between the expected observed probability distribution $P_{N,i}$ and the true one $P_N$ is quantified by
$$
\chi_i^2\equiv \sum_{N=0}^{N_{max}}\left ( \frac{P_N}{P_{N,i}}-1 \right )^2.
$$
It is therefore possible to know the evolution of the cost function $\chi^2$ with respect to the steps $i$. 

In fact it has been shown by \citet{S&P04} that it converges toward a
constant value which corresponds to the best evaluation of the
probability density function $p$ given the observed probability
distribution $P_N$. Since these authors have shown that this
convergence occurs after around $30$ iterations. We did our own
convergence tests which have shown that adopting a value of $30$
iterations is enough. However, it happens that the evolution of the
$\chi^2$ is not always monotonic. In practice, we store the
$\chi^2$ result of each step and we look for the step for which the $\chi^2$ is minimum, i.e. $p(\delta)=p_{i_{min}}(\delta)$. As an initial guess we set that the discret CPDF is equal to the continuous one ($p_0(\delta)=p$).

\subsection{The Skewed Log-Normal}
\label{SLN}

This is a parametric method where the shape of the probability density depends on a given number of parameters, in this case the probability density function is assumed to be well described by a Skewed Log-Normal \citep{colombi94} distribution. It is derived from the Log-Normal distribution \citep{C&J91} but it is more flexible. It is indeed built upon an Edgeworth expansion; be the stochastic field $\Phi \equiv \ln(1+\delta)$, following a Normal distribution then the density contrast $\delta$ follows instead a Log-Normal distribution. In the case of the Skewed Log-Normal (SLN) density function, the field $\Phi$ follows an Edgeworth expanded Normal distribution

\begin{equation}
P_\Phi (\Phi)\equiv \left\{ 1 + \frac{\langle\nu^3\rangle_c}{6} H_3(\nu) + \frac{\langle\nu^4\rangle_c}{24} H_4(\nu) + \frac{5}{72}\langle\nu^3\rangle_c^2 H_6(\nu) \right\}\frac{G(\nu)}{\sigma_\Phi},
\label{edgeworth}
\end{equation} 
where $\nu\equiv \frac{\Phi-\mu_\phi}{\sigma_{\Phi}}$, $G$ is the central reduced Normal distribution $G(\nu)\equiv\frac{e^{-\frac{\nu^2}{2}}}{\sqrt{2\pi}}$ and $\langle\nu^n\rangle_c$ denotes the cumulant expectation value of $\nu$. As a result, the SLN is parameterized by the four parameters $\mu_\Phi$, $\sigma_\Phi$, $\langle\nu^3\rangle_c$, and $\langle\nu^4\rangle_c$ which are related, respectively to the mean, the dispersion, the skewness and the kurtosis of the stochastic variable $\Phi$. They can all be expressed in terms of cumulants $\langle\Phi^n\rangle_c$ of order $n$ of the weakly non-Gaussian field $\Phi$. In \citet{S&P04} they use a best fit approach and determine these parameters by minimizing the difference between the measured counting probability $P_N$ and the one obtained from

\begin{eqnarray}
P_N^{th}= & \int_{-1}^{\infty}K\left[N, \bar N(1+\delta)\right]  P_\Phi\left[\ln(1+\delta), \mu_\Phi, \sigma_\Phi^2, \langle\Phi^3\rangle_c,\langle\Phi^4\rangle_c\right]  \nonumber \\
& \times \dif\ln(1+\delta). 
\label{pth}
\end{eqnarray}
 However, this requires us to perform the integral (Eq.~\ref{pth}) in a four dimensional parameter space which is numerically expensive. 

In the present paper we use an alternative implementation which is computationally more efficient. Instead of trying to maximize the likelihood of the model given the observations, we rather use the observations to predict the parameters of the SLN. To do so we use the property of the local Poisson sampling \citep{B&M12}; the factorial moments $\langle(N)_f^n\rangle$ of the discrete counts are equal to the moments of the underlying continuous distribution $\langle\Lambda^n\rangle$. Since the transformation between the density contrast $\delta$ and the Edgeworth expanded field $\Phi$ is local and deterministic, it is possible to find a relation between the moments $\langle\Lambda^n\rangle$ and the cumulants $\langle\Phi^n\rangle_c$. 

By definition, the moments of the positive continuous field $\Lambda$ are given by 
$$
\langle\Lambda^n\rangle\equiv \int_0^\infty\Lambda^n P(\Lambda)\dif\Lambda,
$$
then for a local deterministic transformation the conservation of probability imposes $P(\Lambda)\dif\Lambda=P_\Phi(\Phi)\dif\Phi$, it follows that the moments of $\Lambda$ can be recast in terms of $\Phi$
$$
\langle\Lambda^n\rangle={\bar\Lambda}^n\int_0^\infty e^{n\Phi} P_\Phi(\Phi)\dif\Phi.
$$
In the right hand side one can recognize the definition of the moment generating function $\mm_\Phi(t)\equiv \langle e^{t\Phi}\rangle$ we therefore obtain that 

\begin{equation}
\mm_\Phi(t=n)=\frac{\langle\Lambda^n\rangle}{{\bar\Lambda}^n}\equiv A_n.
\label{cumumom}
\end{equation}
This equation allows us to link the moment of $\Lambda$ to the cumulants of $\Phi$ via the moment generating function $\mm_\Phi$.

Moreover, since the probability density $P_\phi$ is the product of a sum of Hermite polynomials with a Gaussian function it is straightforward to compute the explicit expression of the moment generating function we obtain

\begin{equation}
\mm_\Phi(t)=\left\{ 1 + \langle\Phi^3\rangle_c\frac{t^3}{6} + \langle\Phi^4\rangle_c\frac{t^4}{24} +  \langle\Phi^3\rangle_c^2\frac{5}{72}t^6 \right\} e^{t\mu_\Phi + t^2\frac{\sigma_\Phi^2}{2}}.
\label{mphi}
\end{equation}
As a matter of fact, Eq.~(\ref{mphi}) and Eq.~(\ref{cumumom}) together allow to set up a system of four equations, for $n={1,2,3,4}$ it reads

\begin{equation}
Y^{n^2}X^nB_n=A_n,
\label{system}
\end{equation}
where $Y\equiv e^{\frac{\sigma_\Phi^2}{2}}$, $X\equiv e^{\mu_\Phi}$ and $B_n\equiv \mm_\Phi(t=n, \mu_\Phi=0,\sigma_\Phi=0)$. In the system of equations (Eq.~\ref{system}) the right hand side is given by observations and the left hand side depends on the cumulants $\mu_\Phi$, $\sigma_\Phi^2$, $\langle\Phi\rangle_c^3$ and $\langle\Phi\rangle_c^4$ parameterized in terms of $X$, $Y$, $x\equiv \langle\Phi\rangle_c^3 $ and $y\equiv \langle\Phi\rangle_c^4 $. In appendix (\ref{systemequation}), we detail the procedure to solve this non-linear system of equations.

We therefore, get the values of the four parameters of the SLN by simply measuring the moments of the counting variable $N$ up to the fourth order.

\subsection{The Gamma expansion}

The Gamma expansion method follows the same idea as described in \S \ref{SLN} but it uses a Gamma distribution instead of a Gaussian one. It uses the orthogonality properties of the Laguerre polynomials in order to modify the moments of the Gamma PDF. Such an expansion has been investigated in \citet{GF&E00} where they compared it to the Edgeworth expansion in order to model the one-point PDF of the matter density field. Then it has been further extended, in a more general context, to multi-point distributions by \citet{M&D}.

As mentioned above the Gamma expansion requires the use of the Gamma distribution $\phi_G$ defined as

\begin{equation}
\phi_G(u)\equiv \frac{u^{k-1}}{\theta\Gamma(k)}e^{-u},
\label{defgam}
\end{equation}
where $\Gamma$ is the Gamma function (for an integer $n$, $\Gamma(n+1)=n!$, $\theta$ and $k$ are two parameters which are related to the two first moments of the PDF. If the galaxy probability density function is well described by a Gamma expansion at order $n$ then it can be formally written as

\begin{equation}
P(\Lambda)=\phi_G(u)f_n^{(k-1)}(u),
\label{gamexp}
\end{equation}
where by definition $u\equiv \frac{\Lambda}{\theta}$, $k=\frac{{\bar \Lambda}^2}{\sigma_\Lambda^2}$, $\theta\equiv \frac{\bar \Lambda}{k}=\frac{\sigma_\Lambda^2}{\bar \Lambda}$. The function $f_n^{(k-1)}$ represents the expansion aiming at tuning the moments of the Gamma distribution; note that the exponent $(k-1)$ is not the derivative of order $k-1$. Since this expansion is built upon the orthogonal properties of products of Laguerre polynomials with the Gamma distribution, the function $f_n^{(k-1)}$ is given by the sum

\begin{equation}
f_n^{(k-1)}(x)\equiv \sum_{i=0}^nc_iL_i^{(k-1)}(x),
\label{fn}
\end{equation}
where $L_i^{(k-1)}$ are the generalized Laguerre polynomials of order $i$ and the coefficients $c_i$ represent the coefficients of the Gamma expansion and therefore depend on the moments of the galaxy field $\Lambda$

\begin{equation}
c_n\equiv \sum_{i=0}^n {n \choose i} \frac{\Gamma(k)}{\Gamma(k+i)}(-1)^i\frac{\langle\Lambda^i\rangle}{\theta^i}.
\label{cn}
\end{equation}
The main interrest of the Gamma expansion with respect to the SLN is that the coefficients of the expansion are directly related to the moments of the distribution we want to model, i.e. it is not necessary to solve a complicated non-linear system of equations nor to perform a Likelihood estimation of the coefficients. Moreover, it can be easily performed at higher order to describe as best as possible the underlying probability density function of galaxies. 

Another advantage of describing the galaxy field $\Lambda$ by a Gamma
expansion probability density function is that the corresponding
observed $P_N$ can be expressed analytically, which is not the case
for the SLN which must be integrated numerically. 

In Appendix (\ref{genfunc}) we demonstrate the previous statement, it follows that the CPDF $P_N$ can be calculated from

\begin{equation}
P_N=\frac{(-\theta)^N}{N!}\sum_{i=0}^nc_i\frac{\Gamma(i+k)}{\Gamma(k)}h_i^{(N)}(\theta),
\label{pngam}
\end{equation}
where $h_i\equiv \frac{1}{i!}\frac{\theta^i}{(1+\theta)^{i+k}}$ and in this case we use the notation $h_i^{(N)}=\frac{\dif^N h_i}{\dif\gamma^N}$. The successive derivatives of $h_i$ can be obtained from the recursive relation
$$
h_i^{(N)}(\theta)=\frac{(i)_f^N}{\theta^N}h_i(\theta) - \sum_{m=1}^{N} {N \choose m} \frac{(i+k)_f^m}{(1+\theta)^m}h_i^{(N-m)}(\theta). 
$$
In addition to the fact that having the possibility of computing the corresponding observed $P_N$ without requiring an infinite integral for each number $N$ is computationally more efficient, it is also practical to have the analytical calculation for some peculiar values of the $k$ parameter of the distribution. In fact, when $k$ is lower than $1$ which occurs on small scales ($4h^{-1}$Mpc), the probability density function goes to infinity when $\Lambda$ goes to $0$ (although the distribution is still well defined).  In particular, this numerical divergence would induce large numerical uncertainties in the computation of the void probability $P_0$. In addition, one can see that for the void probability we have the simple relation 

\begin{equation}
P_0=\sum_{i=0}^{n}c_i\frac{\Gamma(k+i)}{\Gamma(k)}h_i(\theta),
\label{po}
\end{equation}
which can be used to recover the true void probability in VIPERS. 

\section{Application of the methods on a synthetic galaxy distribution}
\label{synthetic}

In this section we analyse a suite of synthetic galaxy
distributions generated from $20$ realizations of a
Gaussian stochastic field. The full process involved in generating these bench-mark catalogues is detailed in Appendix \ref{fake}. Each comoving volume has a cubical geometry of size $500h^{-1}$Mpc. 
We generate the galaxies by discretizing the density field according to the sampling conditional probability $P[N|\Lambda]$ which we assume to be a Poisson distribution with mean $\Lambda$. 
In this way we know the true underlying galaxy density contrast $\delta$. We can therefore perform a fair comparison between the methods introduced in \S 3. 

In order to avoid the effect of the grid ($0.95h^{-1}$Mpc) we smooth both the density field and the discrete field using a spherical Top-Hat filter of radius $R=8h^{-1}$Mpc. We apply the three methods mentioned in \S 3 and compare the reconstructed probability density function to the expected one obtained directly from the density field $\delta$. 

\begin{figure}
\centering \includegraphics[width=65mm,angle=0]{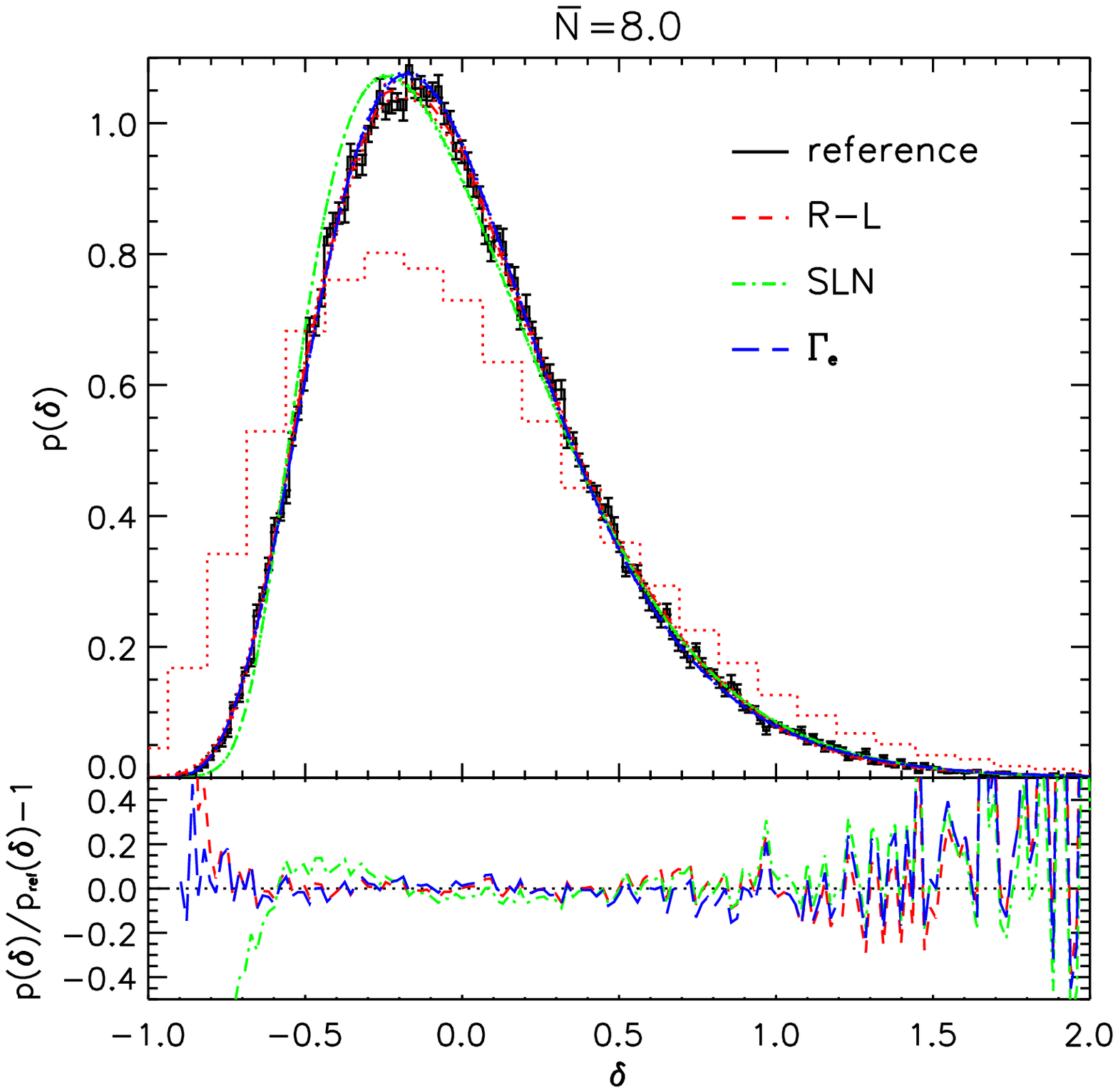}
\includegraphics[width=65mm,angle=0]{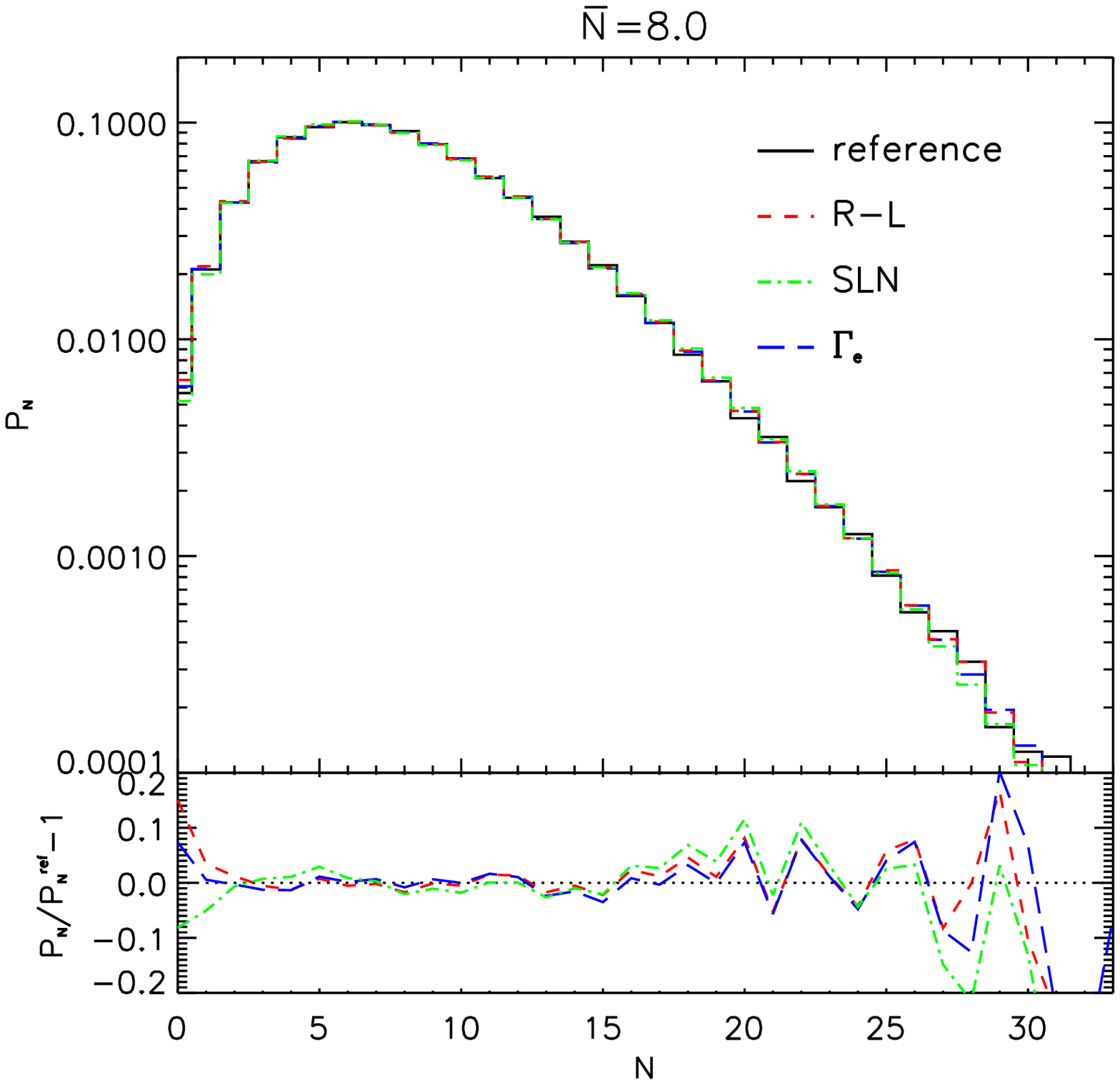}
\caption{ \textit{Upper}: The black histogram with error bars shows the true underlying probability density function (referred to as reference in the inset) compared to the reconstruction obtained with the R-L (red dashed line), the SLN (green dot-dashed line), and the $\Gamma_\mathrm{e}$ (blue long dashed line) methods. The red dotted histogram shows the PDF used as the initial guess for the R-L method and the colored dotted lines around each method line represent the dispersion of the reconstruction among the $20$ fake galaxy catalogues. We also display the relative difference of the result obtained from each method with respect to the true PDF. \textit{Lower}: The black histogram with error bars shows the observed probability density function (referred to as reference in the inset) compared to the reconstruction obtained with the R-L (red dashed line), the SLN (green dot-dashed line), and the $\Gamma_\mathrm{e}$ (blue long dashed line) methods. We also display the relative difference of the result obtained from each method with respect to the observed $P_N$.
}
\label{pd_compare}
\end{figure}

The discrete distribution of points contains an average number of object per cell $\bar N=8$ which is the one expected according to our sampling process. The corresponding $P_N$ is given by the black histogram in the lower panel of Fig.~(\ref{pd_compare}), from this measurement we apply the three methods R-L, SLN and $\Gamma_\mathrm{e}$ and obtain an estimation of the probability density function corresponding to each method. In the upper panel of Fig.~(\ref{pd_compare}) we compare the performance of the three methods in recovering the true probability density function (black histogram referred to as reference in the inset). 
Note that, for this test case, we use a Gamma expansion at order $4$ in order to be coherent with the order of the expansion of the Skewed Log-Normal. We have also represented the probability density function estimated when neglecting the shot noise (red dotted line), which is used as the initial guess in the case of the R-L method. 

From the top panel of Fig.~(\ref{pd_compare}) we can  conclude that the three methods perform reasonably well. It seems that the $\Gamma_\mathrm{e}$ method reproduces better the density distribution of under-dense regions ($\delta \sim -1$) but this is expected in the sense that the distribution used to generate the synthetic catalogues is a Gamma distribution (see Appendix \ref{fake}). Although, it is not obvious because the scale on which the density field has been set up is one order of magnitude smaller than the scale of the reconstruction $R=8h^{-1}$Mpc. 



The performance of the three methods is also represented in the bottom panel of Fig.~(\ref{pd_compare}), in which we compare the expected observed $P_N$ from each method to the true one. One can see that they all agree at the $15$\% level, hence it is not possible to conclude that one is better than an other. This was actually expected, from the comparison on the underlying density field (Fig.~\ref{pd_compare}). On the contrary if one of the methods would not agree with the PDF then we would expect also a disagreement on the observed CPDF (see \S \ref{PDR1}).

\begin{figure}
\centering 
\includegraphics[width=65mm,angle=0]{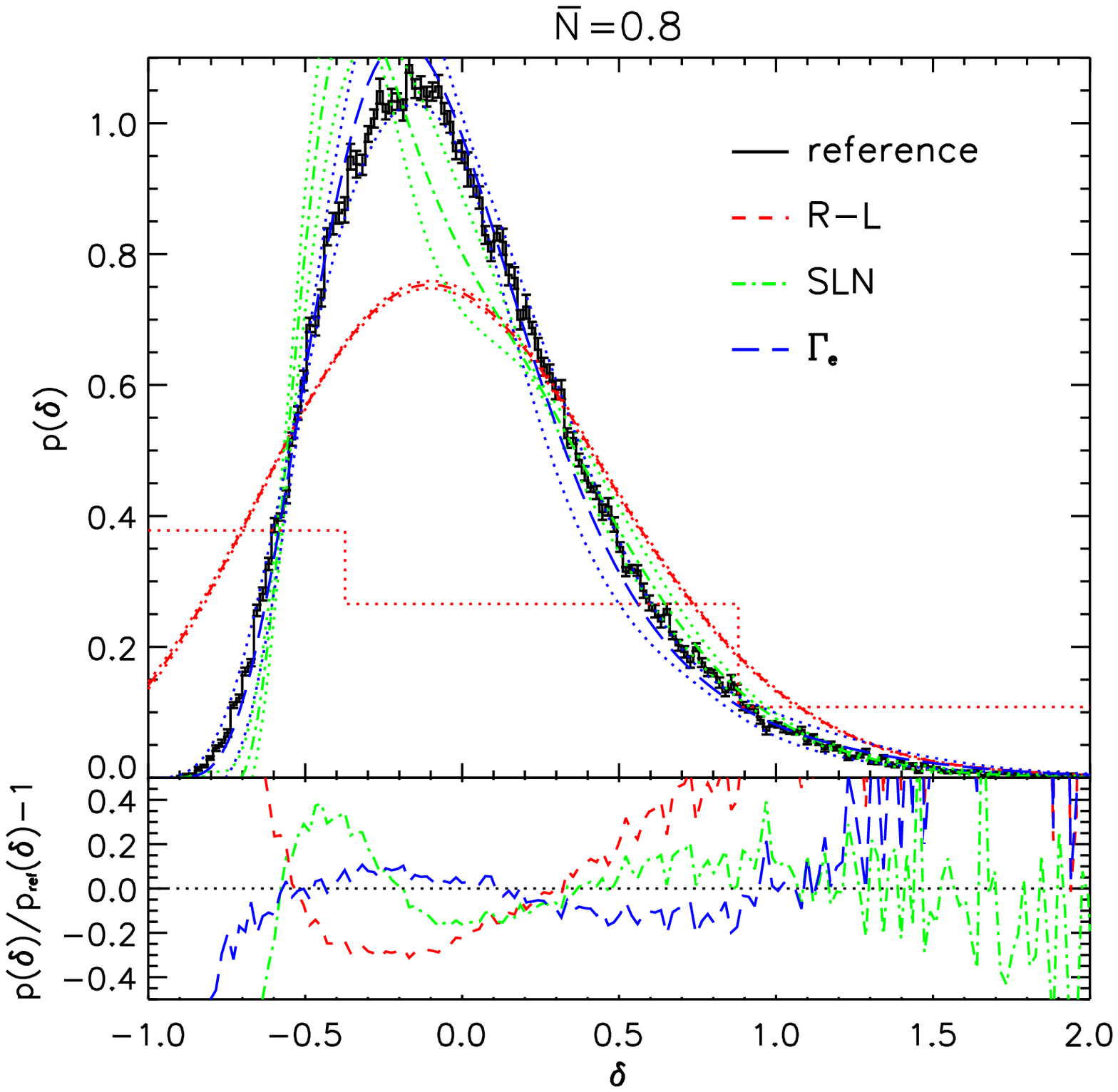}
\includegraphics[width=65mm,angle=0]{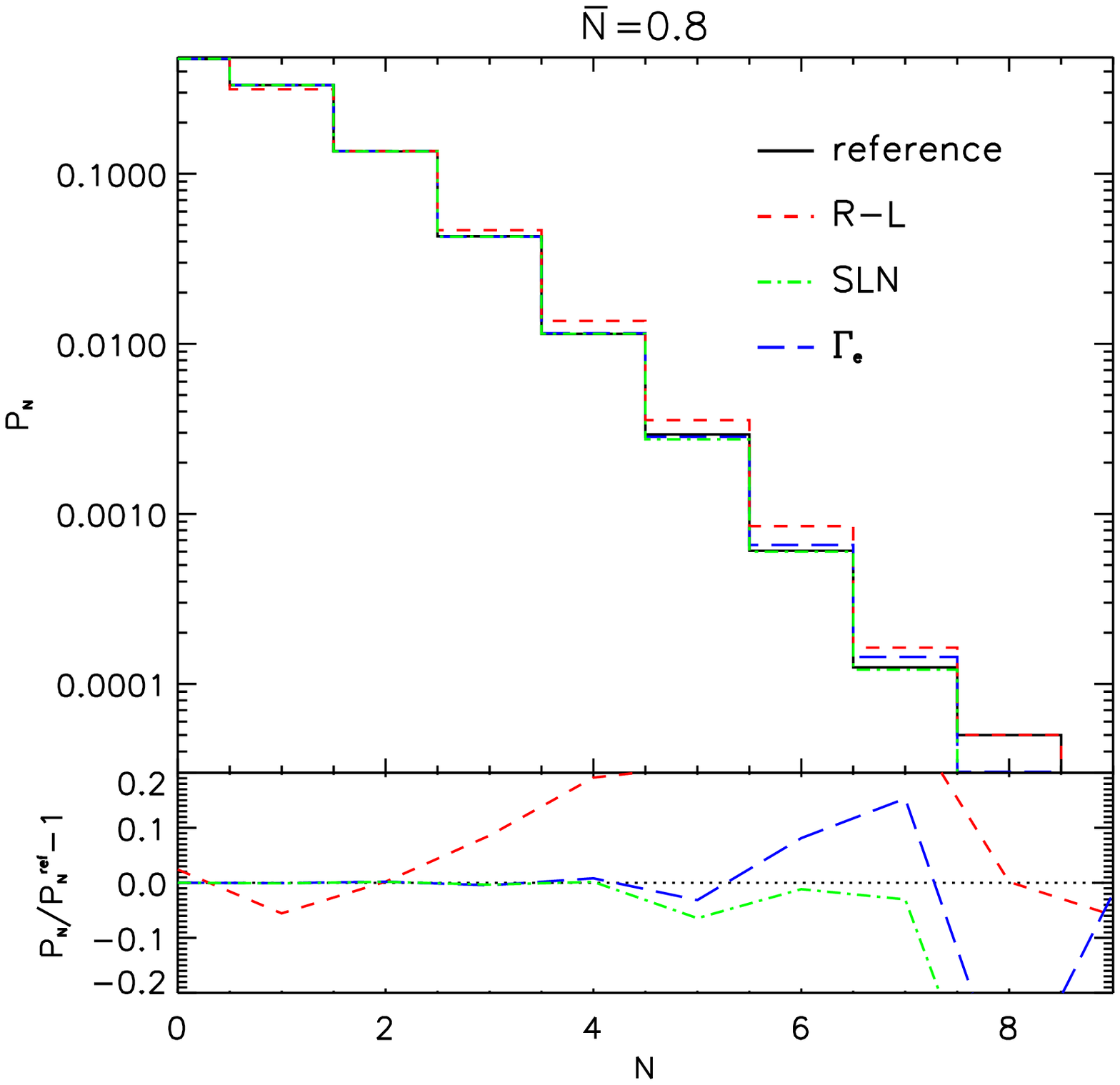}
\caption{ Same as in Fig.~(\ref{pd_compare}) but we use only 10\% of the galaxies contained in the fake galaxy catalogues as a result the average number of galaxy per cell drops from $\bar N=8$ to $\bar N=0.8$.
}
\label{pn_compare}
\end{figure}

In the following part we investigate the sensitivity of the three methods with respect to the shot noise. In fact, as shown in Fig.~(\ref{nbar}), in most of the sub-samples of VIPERS PDR-1 we will work with a high shot noise level ($\bar N \le 1$).  We therefore randomly under-sample the fake galaxy distribution by keeping only $10$\% of the total number of object contained in each comoving volume. This process gives an average number per cell of $0.8$, which is more representative in the context of the application of the reconstruction method. We perform the same comparison as in the ideal case ($\bar N \simeq 8$) and found that the R-L method appears to be highly sensitive to the shot noise. In fact if the mean number of object per cell is too few then the output of the method depends too much on the initial guess. It follows that, if it is too far from the true PDF the process does not converge  (see top panel of Fig. \ref{pn_compare}) and the corresponding $P_N$ does not match the observed one (see bottom panel of Fig. \ref{pn_compare}). Note that we explicitly checked this effect by increasing the number of iterations from $30$ to $200$.  While in the case of both, the SLN and the Gamma expansion, one can see in Fig.~(\ref{pn_compare}), the output probability density function is in agreement (with a larger scatter) with the one obtained in the $\bar N \simeq 8$ case. This means that the sensitivity regarding to the shot noise is much smaller when considering parametric methods. 

%

Considered the sensitivity of the R-L method to the initial guess, knowing that the average number of galaxies per cell can be lower than unity and finally taking into account computational time, we shall continue our analysis only using the two parametric methods SLN and $\Gamma_\mathrm{e}$. In the following, we will compare them using more realistic mock catalogues but for which we don't know apriori the true underlying PDF.

\section{Performances in realistic conditions}
\label{realistic_performances}

In this section we discuss how observational effects have been
accounted for in our analysis and test the robustness of the reconstruction methods SLN and Gamma expansion.
For this purpose we use a suite of mock catalogues created from the Millenium simulation, they are also used in the analysis performed by \citet{diPorto}.

We shall compare the reconstruction methods between two catalogues, namely REFERENCE and MOCK. The reference is a galaxy catalogue obtained from semi-analytical models. We simulate the redshift errors of VIPERS PDR-1 by perturbing the redshift (including distortions due to peculiar motions) with a Normally distributed error with \textit{rms} $0.00047(1+z)$. Each MOCK catalogue is built from the corresponding REFERENCE catalogue by applying the same observational strategy \citep{delaTorre} which is applied on VIPERS PDR-1; spectroscopic targets are selected from the REFERENCE catalogue by applying the 
slit-positioning algorithm \citep[SPOC,][]{Bottini05} with the same setting as for the PDR-1. This
allows us to reproduce the VIPERS footprint on the sky, the
small-scale angular incompleteness due to spectra collisions and the variation of target sampling rate across
the fields. Finally, we deplete each quadrant to reproduce the effect
of the survey success rate \citep[SSR, see ][]{delaTorre}.  In this way, we end up with 50 realistic mock catalogues (named MOCK hereafter), which simulate the detailed survey completeness function and observational biases of VIPERS in the W1 and W4 fields. 

In order to perform a similar analysis as the one we aim at doing for VIPERS PDR-1, we construct sub-samples of galaxies selected according to their absolute magnitude $M_B$ in B-band; we take all objects brighter than a given luminosity. We list those samples in Tab.~(\ref{lumimock}), we have in total $6$ galaxy samples. The highest luminosity cut (M$_B-5\log(h)<19.72-z$) allows us to follow a single population of galaxies at three cosmic epocs.

%
\begin{table}
\centering
  \label{lumimock}
\caption{List of the magnitude selected objects (in B-band) in the mock catalogues}
  \begin{tabular}{c c c}
 \hline
 $z_{\mathrm{min}}$ & $z_{\mathrm{min}}$ & luminosity  \\
 &   & M$_B - 5\log(h) <$ \\
     \hline 
$0.5$	&	$0.7$	&	$-18.42-z$		 \\ 
$0.5$	&    $0.7$	&	$-19.12-z$	         \\ 
$0.5$	&	$0.7$	&	$-19.72-z$	         \\ 
\hline
$0.7$	&	$0.9$	&	$-19.12-z$		 \\ 
$0.7$	&    $0.9$	&	$-19.72-z$	         \\ 
\hline
$0.9$	&	$1.1$	&	$-19.72-z$		 \\ 
\hline
   \end{tabular}

\end{table}

%

\begin{figure}
\includegraphics[width=90mm,angle=0]{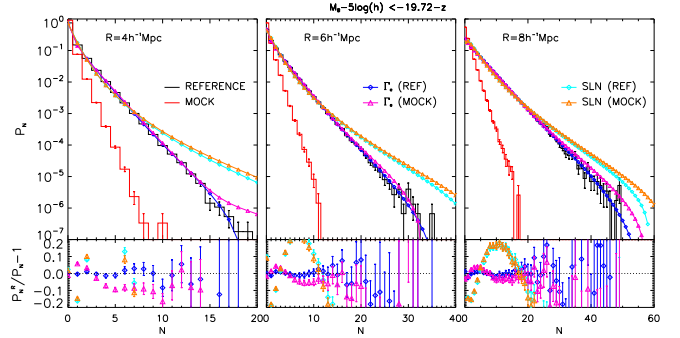}
\caption{Comparison between the SLN and $\Gamma_\mathrm{e}$ methods at $0.9<z<1.1$. Each panel corresponds to a cell radius $R$ of $4$, $6$ and $8h^{-1}$Mpc from the left to the right. \textit{Top}: The red histogram shows the observed PDF in the MOCK catalogues while the black histogram displays the PDF extracted from the REFERENCE catalogues. The blue diamonds with lines and the magenta triangles show, respectively, the $\Gamma_\mathrm{e}$ expansion performed in the REFERENCE and MOCK catalogues. On the other hand, the cyan diamonds with lines and the orange triangles show, respectively, the SLN expansion performed in the REFERENCE and MOCK catalogues. \textit{Bottom}: Relative deviation of the $\Gamma_\mathrm{e}$ and SLN expansions applied both on the REFERENCE and MOCK catalogues with respect to the PDF of the REFERENCE catalogues.
}
\label{figz9}
\end{figure}
%
\begin{figure}
\centerline{\includegraphics[width=90mm,angle=0]{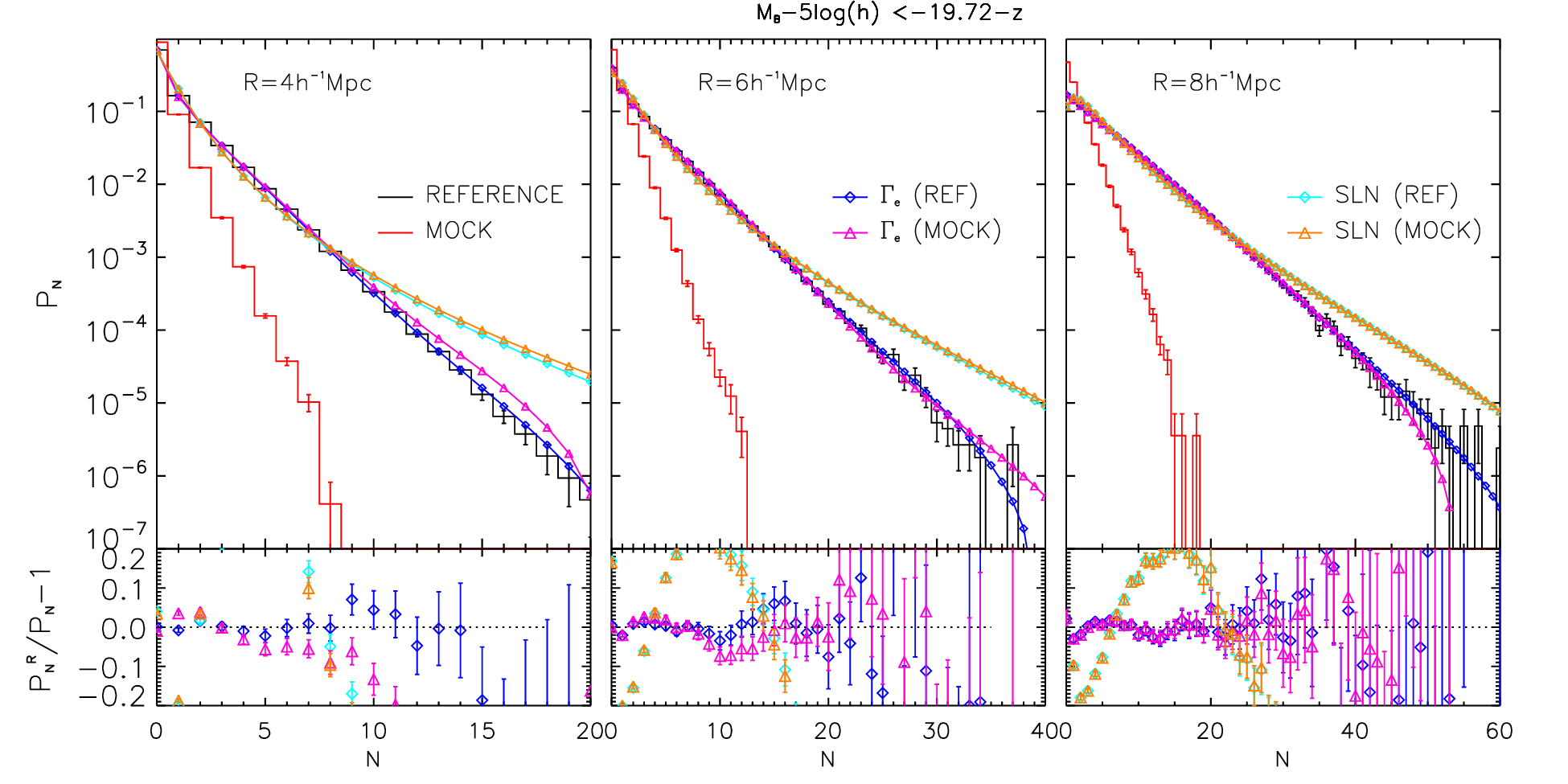} }
\caption{Comparison between the SLN and $\Gamma_\mathrm{e}$ methods at $0.7<z<0.9$. Each panel corresponds to a cell radius $R$ of $4$, $6$ and $8h^{-1}$Mpc from the left to the right. 
}
\label{figz7}
\end{figure}
%
\begin{figure}
\includegraphics[width=90mm,angle=0]{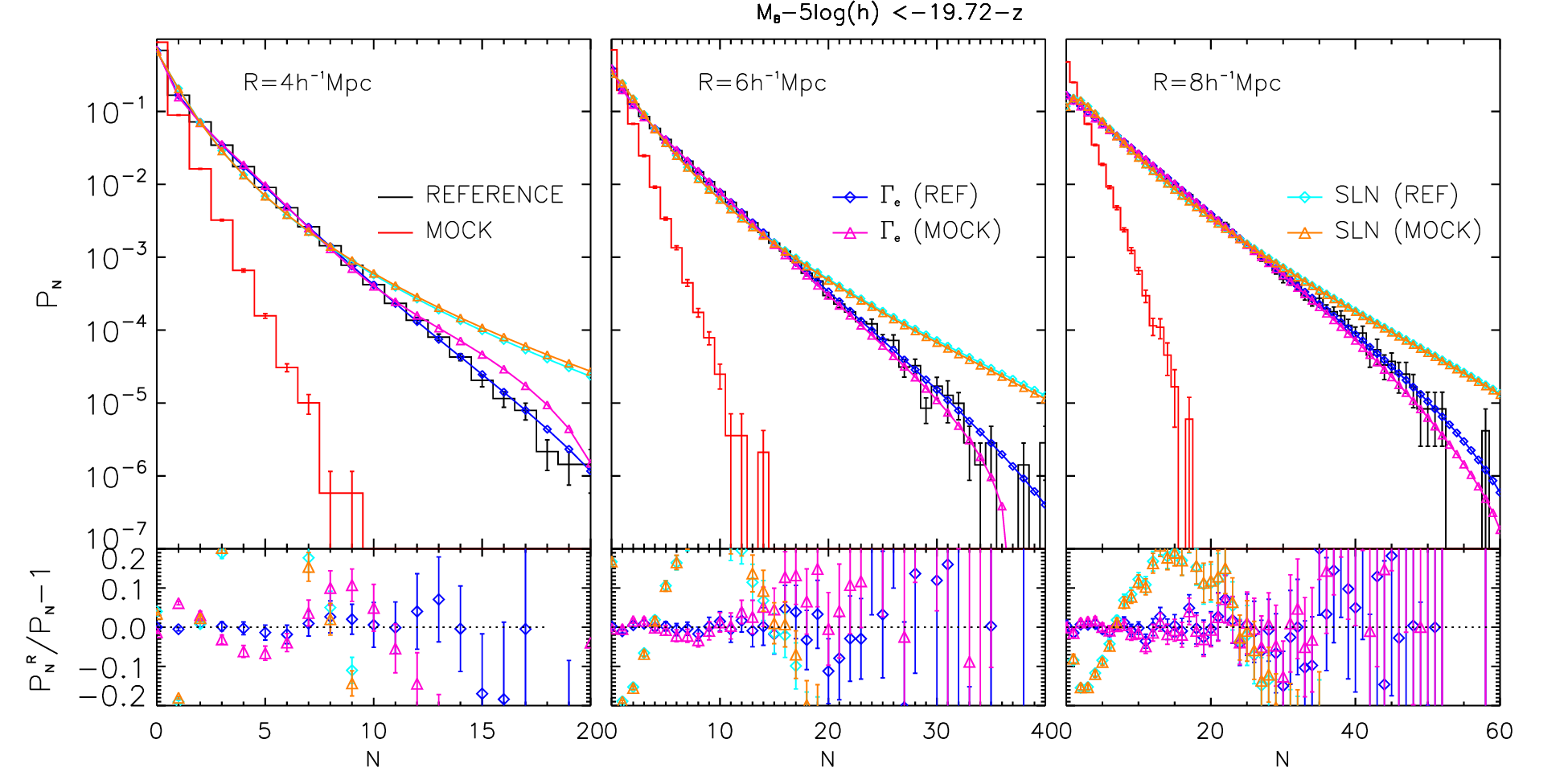}
\caption{Comparison between the SLN and $\Gamma_\mathrm{e}$ methods at $0.5<z<0.7$. Each panel corresponds to a cell radius $R$ of $4$, $6$ and $8h^{-1}$Mpc from the left to the right. 
}
\label{figz5}
\end{figure}
%
\begin{figure}
\includegraphics[width=90mm,angle=0]{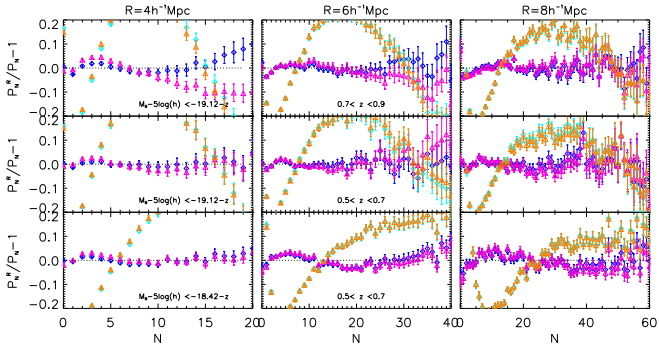}
\caption{Comparison between the SLN and $\Gamma_\mathrm{e}$ methods. Each column corresponds to a cell radius $R$ of $4$, $6$ and $8h^{-1}$Mpc from the left to the right, and each row corresponds to a combination of redshift and magnitude cut. 
}
\label{figzm}
\end{figure}

In Fig.~\ref{figz9}, \ref{figz7} and \ref{figz5} we show the reconstruction performances for the SLN and the $\Gamma_\mathrm{e}$ method. We consider the same population ($M_b-5\log h +z < -19.72$) but in three redshift bins, $0.9<z<1.1$, $0.7<z<0.9$ and $0.5<z<0.7$. In order to test the stability of the methods we perform the reconstruction at three smoothing scales, $R=4$, $6$ and $8h^{-1}$Mpc. The comparison is done as follows, on one hand we estimate the true $P_N$ from the REFERENCE catalogue (before applying the observational selection) and we perform the reconstruction on it, in this way we can test the intrinsic biases due to the assumed parametric method (SLN or $\Gamma_\mathrm{e}$). On the other hand, we estimate the observed $P_M$ in the MOCK catalogues, from which we perform the reconstruction to verify if we recover the expected $P_N$ from the REFERENCE catalogue. 

Inspecting Fig.~(\ref{figz9}) we can first see  that the intrinsic error due to the specific modeling of the methods is much larger for the SLN (cyan diamonds compared to the black histogram) than for the $\Gamma_\mathrm{e}$ (magenta diamonds compared to the black histogram). From the top panels we see that the SLN does not reproduce the tail of the CPDF and from the bottom panel we see that even for low counts it is showing deviations as large as $20$\%. This intrinsic limitation is propagating when performing the reconstruction on the MOCK catalogue (orange triangles compared to the black histogram) while for the $\Gamma_\mathrm{e}$ we see that the agreement is better than $10$\% (magenta triangles compared to the black histogram) in the low count regime and the tail is fairly well reproduced. In the second place, comparing the $\Gamma_\mathrm{e}$ performed on the REFERENCE and the MOCK catalogues (blue diamonds with respect to magenta triangles) one can see the loss of information due to the observational strategy has at most an impact of $10$\% on the reconstructed CPDF which reduces when considering larger cells (less shot noise).

 In general, examination of Fig.~(\ref{figz7}) and (\ref{figz5}) confirms that for the considered galaxy population the same results hold at lower redshifts. However, in particular the reconstruction at $R=4h^{-1}$Mpc can exhibit deviations larger than $20$\%, this is at odds with the fact that the shot noise contribution is expected to be the same for the three redshift bins (magnitude limited). We attribute this larger instability to the fact that not only the  shot noise contribution is higher for $R=4h^{-1}$Mpc but also the volume probed is smaller when decreasing the redshift.  

The performances of the reconstruction for the last three galaxy samples are shown in Fig.(\ref{figzm}) where each row corresponds to a galaxy sample (we only show the residual with respect to the REFERENCE). This last comparison allows to say that the reconstruction instability at $4h^{-1}$Mpc was indeed due to the high level of shot noise.  We can conclude that in the HOD galaxy mock catalogues, the galaxy distribution is more likely to be modelled by a $\Gamma_\mathrm{e}$ instead of an SLN. Finally, for a chosen reconstruction method, the information contained in the MOCK catalogues is enough to be able to reconstruct the CPDF of the REFERENCE catalogue to better than 10\%.

\section{VIPERS PDR-1 data}
\label{PDR1}

In this section we apply the reconstruction method to the VIPERS PDR-1. 
We saw in the previous sections that the SLN and $\Gamma_\mathrm{e}$ methods are sensitive to
the assumption we make about the underlying PDF. In fact, we saw in \S \ref{synthetic}
that if the underlying PDF is close to the chosen model then the reconstruction works. We found in \S \ref{realistic_performances} that the galaxy distribution arising from semi-analytic models is better described by a $\Gamma_\mathrm{e}$ than an SLN distribution. However, in the following we will not take for granted that the same property holds for the galaxies in the PDR-1. 

We want to choose which one of the two distributions (Log-Normal or Gamma) best describes the observed galaxy distribution in VIPERS PDR-1, when no expansion is applied.
Thus, we compare the observed PDF to the one expected from the Poisson sampling of the Log-Normal probability density function (PS-LN) and to the one expected from the Poisson sampling of the Gamma distribution (the so-called Negative Binomial). Error bars are obtained by performing a Jack-knife resampling of $3\times7$ subregions in each fields W1 and W4. 

\begin{figure*}
\centerline{\includegraphics[width=180mm,angle=0]{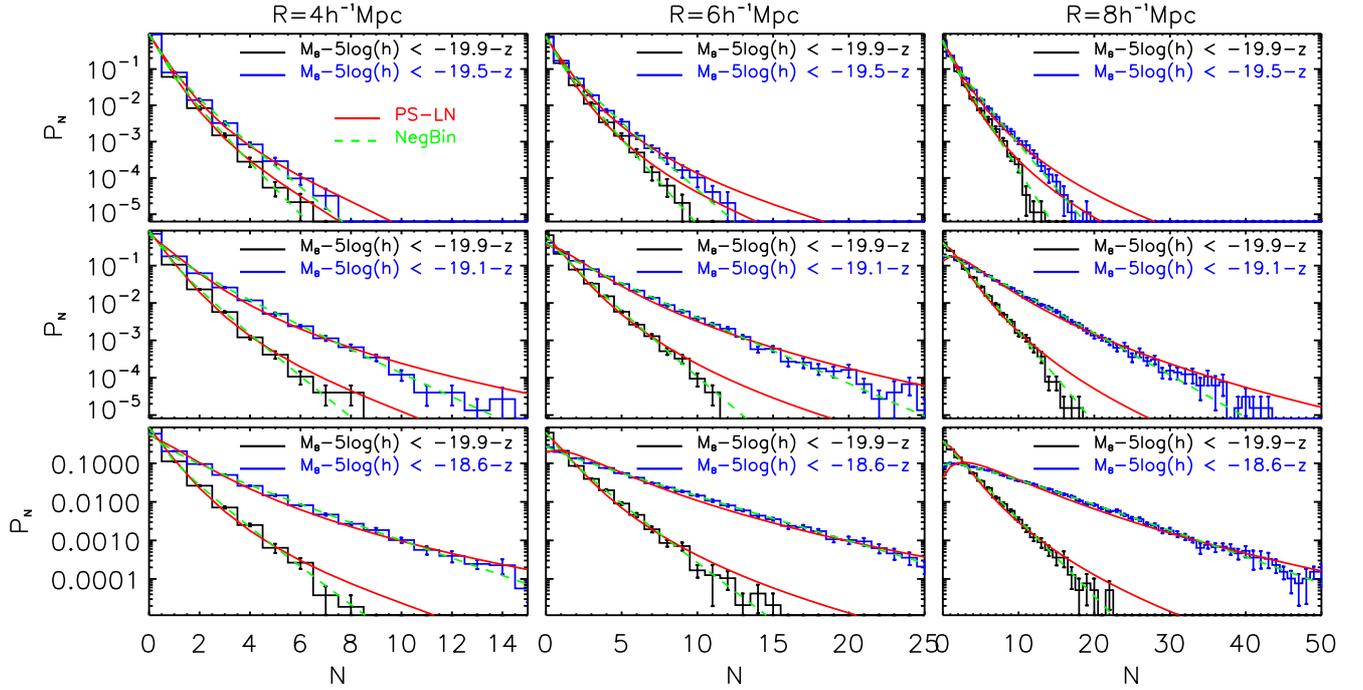} }
\caption{Observed count-in-cell probability distribution function $P_N$ (histograms) from VIPERS PDR-1 for various luminosity cuts (indicated in the inset). Each row corresponds to a redshift bin, from the bottom to the top $0.5<z<0.7$, $0.7<z<0.9$, and $0.9<z<1.1$. Each column corresponds to a cell radius $R={4,6,8}h^{-1}$Mpc from the left to the right. Moreover we added the expected PDF from two models which match the two first moment of the observed distribution; the red solid line shows the prediction for a Poisson sampled Log-Normal (PS-LN) CPDF while the green dashed line displays the Negative Binomial model for the CPDF.
}
\label{pnmask}
\end{figure*}

The SP-LN distribution does not have an analytic expression and must be obtained by numerically integrating Eq.~(\ref{samplingd}) while the Poisson sampling of the Gamma distribution leads to the Negative Binomial distribution defined as

\begin{equation}
P_N=\frac{\theta^N}{N!}\frac{r(r+1)...(r+N-1)}{(1+\theta)^{N+r}},
\label{negbin}
\end{equation}
where $\theta=\frac{\bar N}{r}$ and $r=\frac{\bar N^2}{\sigma_N^2-\bar N}$ to ensure that the first two moments of the Negative Binomial match those of the observed distribution. We show in Fig.~(\ref{pnmask}) the outcome of this comparison, it follows that the Negative Binomial is much closer to the observed PDF than the PS-LN. As a result, the underlying galaxy distribution is more likely to be described by a Gamma distribution than by a Log-Normal. Hence, we only use the Gamma expansion to model the galaxy distribution of VIPERS PDR-1. 

\begin{figure*}
\centerline{\includegraphics[width=180mm,angle=0]{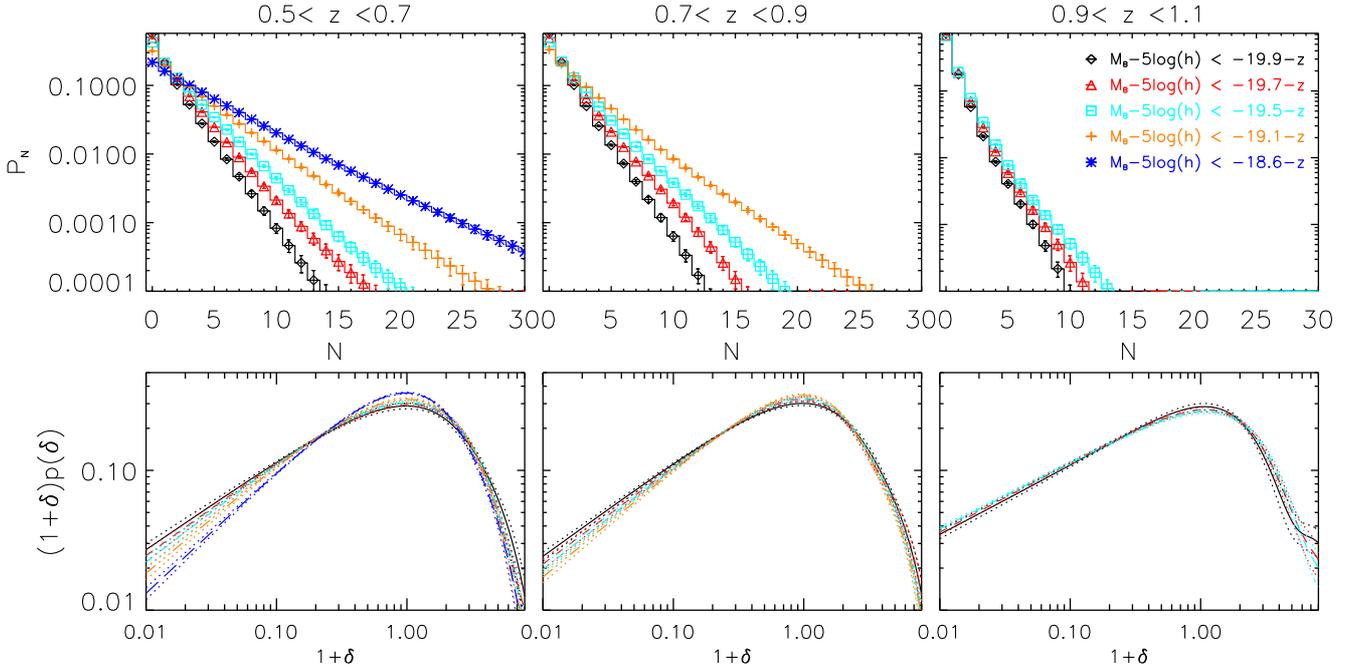} }
\caption{ \textit{Top}: Reconstructed PDF applying the $\Gamma_\mathrm{e}$ method in three redshift bins (from left to right) at the intermediate smoothing scale $R=6h^{-1}$Mpc. \textit{Bottom}: Underlying PDF corresponding to the CPDF in the top panel, for each luminosity cut the 1-sigma uncertainty is represented by the dotted lines.
}
\label{pnpd}
\end{figure*}

Moreover, the use of the Gamma expansion instead of the SLN simplifies substantially  the analysis. In Fig.~(\ref{pnpd}) we provide the reconstructed probability distribution function of VIPERS PDR-1 together with the corresponding underlying probability density function for each redshift bin and luminosity cut. Each panel of Fig.~(\ref{pnpd}) shows how the choice of a particular class of tracers (selected according to their absolute magnitude in B-band) influence the PDF of galaxies. When measuring specific properties of the intrinsic galaxy distribution for each luminosity cut, it is enough to look at the CPDF however, when comparing the distributions with each other it is necessary to take care about the averaged number of objects per cell which varies from sample to sample. As a result it appears more useful to compare the properties of the different galaxy samples using their underlying probability density function which, assuming Poisson sampling, is free from sampling rate variation between different type of tracers. 

For the two first redshift bins, we can see that the probability density function is broadening when selecting more luminous galaxies, this goes in the direction of increasing the linear bias with respect to the matter distribution. However, despite a less significant trend, for the highest redshift bin it seems that it goes in the oposite direction. This trend might be an artifact; indeed by analyzing Fig.~(\ref{nbar}) we see that for all these samples the averaged number of object per cell is between $0.2$-$0.4$ which shows that theses samples could be highly affected by shot noise effects. As a result, specific care should be taken when interpreting those three high redshift samples. 

\begin{figure}
\includegraphics[width=90mm,angle=0]{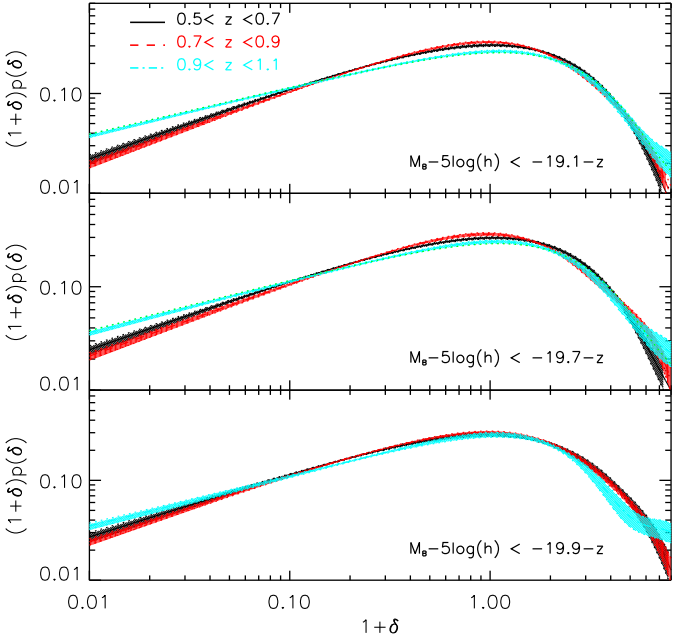} 
\caption{Evolution of three galaxy populations selected according to their luminosity (from bottom to top). On each panel, the black solid, red dashed and, cyand dot-dashed lines represent, respectively, the three redshift bins $0.5<z<0.7$, $0.7<z<0.9$ and, $0.9<z<1.1$.
}
\label{pdfevo}
\end{figure}

In the following we focus on the evolution of the underlying PDF for a particular class of objects on the wide redshift range probed by VIPERS PDR-1. The Fig.~(\ref{pdfevo}) displays the outcome of this study, it shows how the PDF, for three populations (the three highest magnitude cuts), evolves regarding to the redshift at which it is measured.  The three populations (top, middle and bottom panels) exhibit non-monotonic evolution with respect to the redshift. In particular, the more luminous population is showing that the PDF at $0.9<z<1.1$ appears to be systematically different than in the two lower redshift bins. However, we see also that some instabilities are appearing in the reconstruction (see wiggles at high $1+\delta$). This might be due to the fact that we have fewer galaxies in this sample giving rise to a large shot-noise contribution ($\bar N <0.3$). We indeed verified that for the high mass bin and the two other galaxy populations we vary the order of the expansion from $6$ to $4$ the resulting PDF changes by less than $1$-$\sigma$ while for the most luminous population, truncating the expansion at order $4$ only removes the instability without changing significantly the overall behavior of the PDF. This consistency test shows that the radical change in the measured PDF for the highest redshift bin appears to be the true feature.  Probably only the final VIPERS data set will be able to give a robust conclusion.

%

%

%

\begin{table*}
\centering
  \label{coefs}
\caption{Coefficients of the $\Gamma_\mathrm{e}$ expansion which describe the VIPERS PDR-1 data for $R=6h^{-1}$Mpc}
  \begin{tabular}{c c c c c c c c}
 \hline
$z$ &  M$_B - 5\log(h)$ & $k$ & $\theta$ &$c_3$ & $c_4$ & $c_5$ & $c_6$ \\
   \hline
$0.5-0.7$ & $-18.6-z$ &0.87961819   &    4.5053822  &  -0.027583435  &  -0.030026522   & -0.018218867  &  -0.019292756 \\
& $-19.1-z$ &0.78883961   &    3.2677238 &   -0.011759548 &  -0.0041201299  &  0.0076149367 &  -0.0010233871 \\
& $-19.5-z$ & 0.72531432   &    2.2643581&   -0.020667396 &  0.00070338969  &   0.021056193  &-0.00061403852 \\
& $-19.7-z$ &0.64267892   &    1.4068744 &   -0.034276861  &  -0.022797814  &   0.022229339  &   0.023963984 \\
& $-19.9-z$ &0.64267892   &    1.4068744 &  -0.0071341640 &  -0.0072444524 &  -0.0030038079  &  -0.045733910 \\
\hline
$0.7-0.9$ & $-19.1-z$ &0.76911853   &    2.9737929  &  -0.063844766  &  -0.046627985   & -0.032441385   & -0.067589757 \\
& $-19.5-z$ &0.73969794    &   2.0841542  &  -0.032831012 &   -0.032693436  &  -0.028383261  &  -0.064019117 \\
& $-19.7-z$ & 0.70270085  &     1.6638888 &  -0.019063352  &  -0.048572844  &  -0.061832661   & -0.078445546 \\
& $-19.9-z$ & 0.67984433   &    1.2608492 &   0.013646925  &  -0.028325455 &   -0.042087256  &  -0.021113201 \\
\hline
$0.9-1.1$ & $-19.5-z$ &0.47473429   &    1.3138704  &  -0.10794135   &  -0.17074978  &   -0.10267837 & -0.0089188521 \\
& $-19.7-z$ &0.49470455   &    1.0926144  &  -0.075805086  &   -0.16739016   &  -0.13623398  &  -0.019540367 \\
& $-19.9-z$ &0.48382041    &  0.90259279  &  -0.076620326  &   -0.20604275  &   -0.23060122   &  -0.14506575 \\
\hline
 \end{tabular}

\end{table*}

Finally, in Tab.~(\ref{coefs}), we list the relevant coefficients of the Gamma expansion which we measured from the VIPERS PDR-1 at the scale $R=6h^{-1}$Mpc. They can be used in order to model both the CPDF (Eq. \ref{pngam}) and the PDF (Eq. \ref{gamexp}). 

\section{Summary}

The main goal of the present paper is to measure the probability of finding $N$ galaxies falling into a spherical cell randomly placed inside a sparse sampled (i.e. with masked areas or with low sampling rate) spectroscopic survey. Our general approach to this problem has been to use the underlying probability density distribution of the density contrast of galaxies in order to recover the counting probability corrected from sparseness effects. We therefore compared three ways (R-L, SLN and $\Gamma_e$) of measuring the probability density of galaxies classified in two categories; direct and parametric. We found that when the sampling is high ($\bar N \simeq 10$) the direct method (Rychardson-Lucy deconvolution) performs well and avoids putting any prior on the shape of the distribution. On the other hand, we saw that when the sampling is low ($\bar N \simeq 1$) the direct method fails to converge to the true underlying distribution. We thus concluded that, in such cases, the only alternative is to use a parametric method.

We presented two parametric forms aiming at describing the galaxy density distribution, the SLN which is often used in the literature to model the matter distribution and the $\Gamma_\mathrm{e}$. Despite the fact that the two distributions used in this paper have been already investigated in previous works, the approach we propose to estimate their parameters is completely new. Previously, fitting procedures were used in order to estimate them. Here we propose to measure directly the parameters of the distributions from the observations. The method can be applied to both distributions SLN and $\Gamma_\mathrm{e}$ and decreases considerably the computational time of the process. 

Relying on simulated galaxy catalogues of VIPERS PDR1, we tested the reconstruction scheme of the counting probability ($P_N$) under realistic conditions in the case of the SLN and $\Gamma_\mathrm{e}$  expansions. We found, that the reconstruction depends on the choice of the model for the galaxy distribution. However, we have also shown that it is possible to test which distribution better describes the observations.

Using VIPERS PDR1, on the relevant scales investigated in this paper ($R=4,6,8h^{-1}$Mpc), we found that the $\Gamma$ distribution gives a better description of the observed $P_N$ than the one provided by the Log-Normal (see Fig. \ref{pnmask}). We therefore adopted the $\Gamma_\mathrm{e}$ parametric form in order reconstruct the probability density functions of galaxies. From these reconstruction we studied how their PDF changes according to their absolute luminosity in B-band and we also studied their redshift evolution. We found that little evolution has been detected in the two first redshift bins while it seems that the density distribution of the galaxy field is strongly evolving in the last redshift bin. 

We finally used, the measured pdf in order to reconstruct the counting probability (CPDF) one would observe if VIPERS was not masked by gaps between the VIMOS quadrants.

\begin{acknowledgements}

JB acknowledges useful discussions with E. Gazta\~naga.
We acknowledge the crucial
contribution of the ESO staff for the management of service
observations. In particular, we are deeply grateful to M. Hilker for
his constant help and support of this programme. Italian participation
to VIPERS has been funded by INAF through PRIN 2008 and 2010
programmes. JB, LG and BJG acknowledge support of the European Research
Council through the Darklight ERC Advanced Research Grant (\#
291521). OLF acknowledges support of the European Research Council
through the EARLY ERC Advanced Research Grant (\# 268107). AP, KM, and JK have been supported by the National Science Centre (grants UMO-2012/07/B/ST9/04425 and UMO-2013/09/D/ST9/04030), the Polish-Swiss Astro Project (co-financed by a grant from Switzerland, through the Swiss Contribution to the enlarged European Union), the European Associated Laboratory
Astrophysics Poland-France HECOLS and a Japan Society for the
Promotion of Science (JSPS) Postdoctoral Fellowship for Foreign
Researchers (P11802). GDL acknowledges financial support from the
European Research Council under the European Community's Seventh
Framework Programme (FP7/2007-2013)/ERC grant agreement n. 202781. WJP
and RT acknowledge financial support from the European Research
Council under the European Community's Seventh Framework Programme
(FP7/2007-2013)/ERC grant agreement n. 202686. WJP is also grateful
for support from the UK Science and Technology Facilities Council
through the grant ST/I001204/1. EB, FM and LM acknowledge the support
from grants ASI-INAF I/023/12/0 and PRIN MIUR 2010-2011. CM is
grateful for support from specific project funding of the {\it
Institut Universitaire de France} and the LABEX OCEVU.

\end{acknowledgements}

\appendix

\section{Non-linear system}

\label{systemequation}

The problem of this system of equations is that it is non-linear, it is therefore difficult to solve however it can be reduced to a one dimensional equation which can be solved numerically. 

The two first equations ($n=1$ and $n=2$) can be used to express the two first cumulants with respect to the third and fourth order ones

\begin{eqnarray}
\sigma_\Phi^2 & = & \ln(A_2) + \ln\left( \frac{B_1^2}{B_2}\right) \label{sig} \\
\mu_\Phi & = & -\frac{1}{2}\left [\ln(A_2) + \ln\left( \frac{B_1^4}{B_2}\right) \right ] \label{mu}
\end{eqnarray} 
where $B_1$ and $B_2$ are both functions of $x$ and $y$. Then using other combinations of equation one can express a system of two equations for $x$ and $y$ alone

\begin{eqnarray}
B_3^2 & = & a_1B_1^2B_4 \label{eq1}\\
B_3B_1^3 & = & a_2B_2^3,\label{eq2}
\end{eqnarray} 
where $a_1\equiv \frac{A_3^2}{A_4}$ and $a_2\equiv \frac{A_3}{A_2^3}$. In order to solve properly the system we prefer to express it in term of one parameter $\eta\equiv B_2/B_1$, moreover one can see that polynomials $B_1$ to $B_4$ are not independent, as a result 
$$
B_4=d + aB_1 + bB_2 + cB_3,
$$
where $a=96, b=-32, c=\frac{224}{27}, d=-\frac{1925}{27}$ and which can be substituted in Eq.~(\ref{eq1}). Combining Eq.~(\ref{eq1}) and  Eq.~(\ref{eq2}) one obtains a parametric equation for $B_1$

\begin{equation}
(a+b\eta)B_1^3 + (d+cf(\eta))B_1^2 - g(\eta)=0,
\label{b1param}
\end{equation}
which can be solved for each value of the parameter $\eta$ and an independent parametric equation for $B_3$
$$
B_3=f(\eta).
$$
As a result we can find a couple $B_1$,  $B_3$ for each value of the parameter $\eta$, it follows that one can express $x$ and $y$ with respect to $\eta$ and 
given the definition of $\eta$ the possible solution $x$ and $y$ must satisfy the condition 
$$
B_2[x(\eta),y(\eta)]-\eta B_1[x(\eta),y(\eta)]=0,
$$
which gives the possible values of $\eta$ from which one can recover $x$ and $y$. Finally, from Eq.~(\ref{sig}) and Eq.~(\ref{mu}) we can compute the values of $\sigma_\Phi$ and $\mu_\Phi$ corresponding to each couple ($x$, $y$) of solutions. This allows us to select the solution which provides a value of $A_5$ closer to the observed one. 

Once the values of the cumulants $\mu_\Phi$, $\sigma_\Phi^2$, $\langle\Phi^3\rangle_c$ and $\langle\Phi^4\rangle_c$ are known from the process detailed above, we know that the moments of the corresponding $P_N^{th}$ will match those of the observed on up to order $4$. At the end, one can check whether the SLN distribution provides a good match to data by integrating numerically the probability density function convolved with the Poisson kernel $K$ (see Eq. \ref{pkernel}). 

\section{Generating function}
\label{genfunc}

We show that the CPDF associated to a Gamma expanded PDF can be calculated analytically from an expression which depends explicitly on the coefficients $c_i$ of the Gamma expansion.

Be $\mathcal{G}_N$ the generating function associated to the probability distribution $P_N$, it is defined as

\begin{equation}
\mathcal{G}_N(\lambda)\equiv\sum_{i=0}^{\infty}\lambda^NP_N.
\label{gener}
\end{equation}
In case of the Poisson sampling of a Gamma distribution, after some algebra, one can show that it can be expressed with respect to the coefficients of the Gamma expansion as

\begin{equation}
\mathcal{G}_N(\lambda)=\frac{1}{\Gamma(k)}\sum_{i=0}^nc_iF_i(\gamma),
\label{gn}
\end{equation}
where $\gamma\equiv (1-\lambda)\theta$ and 
$$
F_i(\gamma)\equiv \int_{0}^{\infty}x^{k-1}e^{-x}L_i^{(k-1)}(x)e^{-\gamma x}\dif x.
$$
Nevertheless, this integral can be computed using the Laguerre expansion of the exponential 
$$
e^{-\gamma x}=\sum_{i=0}^{\infty}\frac{\gamma^i}{(1+\gamma)^{i+\alpha+1}}L_i^{(\alpha)}(x),
$$
it reads to 

\begin{equation}
F_i(\gamma)=\frac{\gamma^i}{(1+\gamma)^{i+k}}\frac{\Gamma(i+k)}{i!}.
\label{fi}
\end{equation}
The formal expression of the generating function is therefore given by 

\begin{equation}
\mathcal{G}_N(\lambda)=\frac{(1+\gamma)^{-k}}{\Gamma(k)}\sum_{i=0}^nc_i\frac{\Gamma(i+k)}{i!}\left (\frac{\gamma}{1+\gamma} \right ),
\label{expli}
\end{equation}
where we still use $\gamma=(1-\lambda)\theta$. From the explicit expression of the moment generating function (Eq.~\ref{expli}) one can get the probability distribution $P_N$ by iteratively deriving the generating function with respect to $\gamma$ 
$$
P_N\equiv \frac{1}{N!}\left.\frac{\dif^N \mathcal{G}_N(\lambda)}{\dif \lambda^N}\right|_{\lambda=0}=\frac{(-\theta)^N}{N!}\left.\frac{\dif^N \mathcal{G}_N(\gamma)}{\dif \gamma^N}\right|_{\gamma=\theta}.
$$
These derivatives can be calculated explicitly.

\section{Synthetic galaxy catalogues}
\label{fake}

In this Appendix we describe how we generate synthetic galaxy catalogues from Gaussian realizations. The first requirement of these catalogues is that they must be characterized by a known power spectrum and 1-point probability distribution function. The second requirement is that the probability distribution function must be measurable. 

The basic idea is simple, we generate a Gaussian random field in Fourier space (assuming a power spectrum), we inverse Fourier transform it to get its analog in configuration space. We further apply a local transform in order to map the Gaussian field into a stochastic field characterized by the target PDF. The two crucial step of this process are the choice of the input power spectrum and the choice of the local transform. 

Be $\nu$ a stochastic field following a centered ($\langle\nu\rangle=0$) reduced ($\sigma_\nu^2\equiv\langle\nu^2\rangle_c=1$) Gaussian distribution. From a realization of this field, one can generate a non-Gaussian density field $\delta$ by applying a local mapping $L$ between the two, hence

\begin{equation}
\delta = L(\nu).
\label{local}
\end{equation}
The local transform $L$ must be chosen in order to match some target PDF $P_{\delta}$ for the density contrast $\delta$. Assuming that the local transform is a monotonic function which maps the ensemble $]-\infty, +\infty[$ into $]-1, +\infty[$ then, due to the probability conservation $P_{\delta}(\delta)\dif\delta=P_{\nu}(\nu)\dif\nu$, the local transform must verify the following matching

\begin{equation}
\mathcal{C}_\delta[\delta] = \mathcal{C}_\nu[\nu],
\label{matching}
\end{equation} 
where $\mathcal{C}_x$ stands for the cumulative probability distribution function. Be $[a, b]$ the definition assemble of the variable $x$
then its cumulative probability distribution function is defined as $\mathcal{C}_x[x]\equiv \int_{a}^xP_x(x^\prime)\dif x^\prime$, where $P_x$ is the PDF of $x$. By definition a probability density function is positive, it follows that its cumulative is a monotonic function and therefore Eq. (\ref{matching}) can always be inverted, it reads 
$$
\delta=\mathcal{C}_\delta^{-1}\left [ \mathcal{C}_\nu (\nu )  \right],
$$
where the exponent $-1$ stands for the reciprocal function such that $F^{-1}\left[ F(x)\right]=x$. For example, by definition the local mapping $L$ which allows transform a Normal distribution into a Log-Normal distribution is $\delta=e^\nu -1$. Note that depending on the PDF that must be matched this inversion can require a numerical evaluation which can be tabulated. 

Once a local transform is chosen, we need to adress the question of finding the appropriate power spectrum of the Gaussian field $\nu$ which, once locally mapped into the density field $\delta$, will match the expected power spectrum. Following \citet{G&E15}, who considered a log-transform we generalized their result to a generic local transformation. This mapping is not direct in Fourier space while it is in configuration. Writing the two point moment of order two of the density field $\delta$ and assuming the probability conservation leads to 

\begin{equation}
\xi_\delta\equiv \langle\delta_1\delta_2\rangle=\int \mathrm{L}(\nu_1) \mathrm{L}(\nu_2) \mathcal{B}(\nu_1,\nu_2,\xi_\nu)\dif \nu_1\dif \nu_2,
\label{gaussint}
\end{equation}
where $\mathcal{B}$ is a bivariate Gaussian defined as

\begin{equation}
\mathcal{B}(\nu_1,\nu_2,\xi_\nu)\equiv \frac{1}{2\pi|C_\nu|^{1/2}}\mathrm{exp}\left\{ -\frac{1}{2} {\bf \nu}^T C_\nu^{-1} {\bf \nu}\right\}.
\label{biv}
\end{equation}
One can notice that in our case (central reduced Gaussian) the covariance matrix $C_\nu$ takes the simple form
$C_\nu=\left [ 
\begin{array}{cc}
1 & \xi_\nu \\
\xi_\nu & 1
\end{array}
\right]$. Once integrated over the definition domain of $\nu_1$ and $\nu_2$, Eq.~(\ref{gaussint}) provides a mapping between the 2-point correlation function of the Gaussian field $\nu$ and the 2-point correlation function of the density field $\delta$. However, we prefer to rotate the coordinate system before performing the integral (\ref{gaussint}) because in case of high correlation ($\sim 1$) then the gaussian will be comparable with a straight line; most of the sampling of this function will be useless. That's why we look for the rotation allowing to diagonalize the matrix $C_\nu$ and therefore convert ${\bf \nu}$ into a new variable ${\bf x}$.
It follows that 
$$
C_x=\left [ 
\begin{array}{cc}
1 -\xi_\nu & 0\\
0 & 1+\xi_\nu
\end{array}
\right]
$$
and the integral becomes
\begin{equation}
\xi_\delta=\frac{1}{2\pi\sqrt{1-\xi_\nu^2}}\int \mathrm{L}\left(\frac{x_2-x_1}{2}\right) \mathrm{L}\left(\frac{x_2+x_1}{2}\right) e^{-\frac{1}{2} \left( \frac{x_1^2}{\sigma_1^2} + \frac{x_2^2}{\sigma_2^2}   \right)}\dif x_1\dif x_2,
\label{gaussintx}
\end{equation}
where $\sigma_1^2=1-\xi_\nu$ and $\sigma_2^2=1+\xi_\nu$ we can therefore integrate over a bounded domain corresponding to the $-8\sigma_1$, $8\sigma_1$ along the $x_1$ axis and  $-8\sigma_2$, $8\sigma_2$ along the $x_2$ axis. 
An other possibility to perform the integral \ref{gaussint} is to use the Mehler's formula, doing so, one can show that the 2-point correlation of the density field can be expressed as a Taylor expansion on the 2-point correlation function of the $\nu$ field. It reads,

\begin{equation}
\xi_\delta=\lambda(\xi_\nu)\equiv\sum_{n=0}^\infty n!c_n^2\xi_\nu^n,
\label{xiexp}
\end{equation}
where the $c_n$ are the coefficients of the Hermit transform of the local mapping $L(\nu)=\sum_{n=0}^\infty c_nH_n(\nu)$ and they can be calculated using the orthogonal properties of Hermit polynomials 

\begin{equation}
c_n=\frac{1}{n!}\int_{-\infty}^{+\infty}L(\nu)H_n(\nu)P_\nu(\nu)\dif\nu.
\label{cn}
\end{equation}
The latter approach considerably speed up the numerical evaluation of Eq.~(\ref{gaussintx}), it allows to compute the $2$-D integral as a finite sum of $1$-D integrals. It also allows to verify that when the 2-point function of the field $\nu$ is positive then the derivative of $\xi_\delta$ with respect to $\xi_\nu$ is positive. Moreover, from Eq.(\ref{gaussint}) one can see that $\xi_\nu=0$ implies $\xi_\delta=0$. This means that the function which transforms $\xi_\nu$ into $\xi_\delta$ is invertible as long as $\xi_\delta$ is positive. On the other hand we know that the zero-crossing of the 2-point correlation function occurs at very large scales at which one can safely assume that $|\xi_\delta|<<1$ thus by continuity one can truncate the Eq.(\ref{xiexp}) at order one providing a linear relation between $\xi_\delta$ and $\xi_\nu$. As a result, one can take the reciprocal of the function $\lambda$ such that $\xi_\nu=\lambda^{-1}( \xi_\delta )$. 

Once the local transform $L$ and the 2-point correlation mapping $\lambda$ are known, then the input power spectrum of the Gaussian field $\nu$ can be obtained as follow. We choose a power spectrum $P(k)$, in the present case \citet{EH}, for the density field $\delta$, we calculate its corresponding $2$-point correlation function 

\begin{equation}
\xi_\delta=\int P(k)e^{i \vec k\cdot \vec r}\dif^3\vec k.
\label{fback}
\end{equation}
At each scale $r$, one can deduce the 2-point correlation function of the Gaussian field $\xi_\nu=\lambda^{-1}(\xi_\delta)$ and finally using a Fourier transform we obtain the input power spectrum

\begin{equation}
P_{in}(k)=\frac{1}{(2\pi)^3}\int \xi_\nu(r)e^{-i \vec k\cdot \vec r}\dif^3\vec r.
\label{fback}
\end{equation}
Finally, in order to make sure that the PDF target will be reproduced, it is necessary to verify that, once the input power spectrum $P_{in}(k)$ have been set up on regular k-space grid which will be used to generate the Gaussian field, its integral is indeed equal to the expected variance on the size of the mesh. $\hat\sigma_a^2=(\frac{2\pi}{L})^3\sum_{n}P(\vec k_n)$ should be equal to $\sigma_a^2=\int P(k)\dif^3\vec k$. In general, $\sigma_a$ and $\hat\sigma_a$ are not equal, thus we renormalize the target power spectrum by the quantity $S=\hat\sigma_a^2/\sigma_a^2$, $\hat P_{in}(k)=SP_{lin}(k)$.

\begin{figure}
\centerline{ \includegraphics[width=80mm,angle=0]{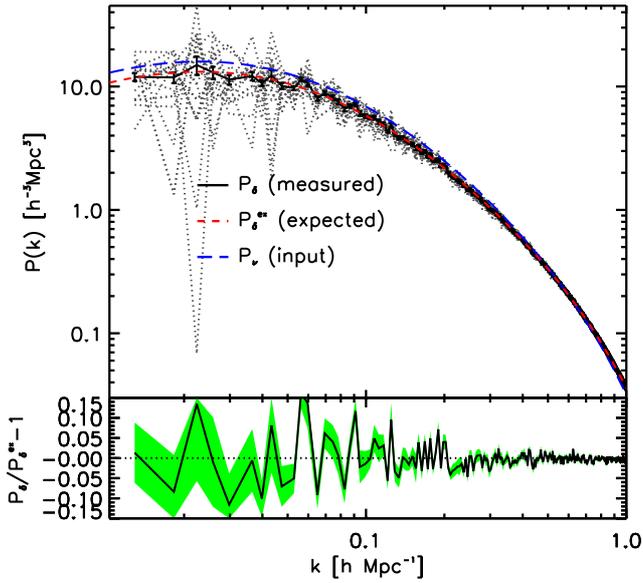} }
\caption{\textit{Upper}: Grey dotted lines show the power spectrum measured in each of the $20$ fake galaxy distributions, the black solid line represent their average and the errors display the dispersion of the measurements. The blue long dashed line displays the input power spectrum used too generate the Gaussian stochastic field $nu$ and the red dashed line shows the corresponding expectation value for the power spectrum of the density contrast $\delta$. \textit{Lower}: Shows the deviation between the measured power spectrum of the $\delta$-field and the expected one.
}
\label{power}
\end{figure}

We generate a Gaussian field (with a flat power spectrum), on a regular mesh of $a=0.95h^{-1}$Mpc and a comoving box of $500^3h^{-3}$Mpc$^3$. We then Fourier transform with an FFT and keep only the phases of the field $\nu_{\vec k}=e^{i\theta(\vec k)}$. We generate at each position $\vec k_n$ the value of the modulus of $\nu_{\vec k}=\sqrt{X_k}e^{i\theta(\vec k)}$, where $X_k=-\hat P_{in}(k)\ln(1-\epsilon)$ and $\epsilon$ is a random number with a uniform probability distribution between $0$ and $1$. We then inverse Fourier transform the field to get a centered reduced Gaussian field. In Fig.~(\ref{power}) we show the input power spectrum of the Gaussian field $\nu$ compared to the one measured using a FFT, and to the one expected from the local transformation applied to the $\nu$ field in order to generate the density field $\delta$.

\end{document}